\documentclass[prb,twocolumn,superscriptaddress,showpacs]{revtex4-1}
\usepackage{graphicx}
\usepackage{epsfig}
\usepackage{amssymb,amsmath,amsfonts,hyperref}
\usepackage{wasysym}
\usepackage{latexsym}
\usepackage{color}

\newcommand{\nn}{\nonumber}

\newcommand{\be}{\begin{eqnarray}}
\newcommand{\ee}{\end{eqnarray}}

\begin{document}
\title{Vacancy-induced spin textures and their interactions in a classical spin liquid}
\author{Arnab Sen}
\affiliation{\small{Max-Planck-Institut f\"{u}r Physik komplexer Systeme, 01187 Dresden, Germany }}

\author{Kedar Damle}
\affiliation{\small{Department of Theoretical Physics, Tata Institute of Fundamental Research, Mumbai 400 005, India.}}

\author{R. Moessner}
\affiliation{\small{Max-Planck-Institut f\"{u}r Physik komplexer Systeme, 01187 Dresden, Germany }}

\begin{abstract}
Motivated by experiments on the archetypal frustrated magnet SrCr$_{9p}$Ga$_{12-9p}$O$_{19}$ (SCGO), we study the classical Heisenberg model on
the pyrochlore slab (Kagom\'e bilayer) lattice with site-dilution $x=1-p$. This allows us to address generic aspects of
the physics of non-magnetic vacancies in a classical
spin liquid. We explicitly demonstrate that the pure ($x=0$) system remains a spin-liquid down to the lowest temperatures, with an unusual {\em non-monotonic} temperature dependence of the susceptibility, which even turns diamagnetic for the apical spins between the two kagome layers. For $x> 0$ but small, the low temperature
magnetic response of the system is most naturally described
in terms of the properties of spatially extended spin textures that 
cloak an ``orphan'' $S=3/2$ Cr$^{3+}$ spin in direct proximity to a pair of
missing sites belonging to the same triangular simplex. In the $T \rightarrow 0$ limit, these orphan-texture
complexes each carry a net magnetization that is exactly half the magnetic moment of an individual spin of the undiluted system. Furthermore, we demonstrate that they  interact via an entropic {\em temperature
dependent} {\em pair-wise exchange interaction} $J_{eff}(T,\vec{r}) \sim T {\mathcal J} (\vec{r} \sqrt{T})$ that has a logarithmic form at short-distances and decays exponentially beyond a thermal correlation length $\xi(T) \sim 1/\sqrt{T}$. The sign of $J_{eff}$ depends on 
whether the two orphan spins belong to the same Kagome layer or not. We provide a detailed analytical account of these
properties using an effective field theory approach specifically tailored for the problem at hand. These results are in quantitative agreement with large-scale Monte Carlo numerics.

\end{abstract}

\date\today
\pacs{75.10.Jm 05.30.Jp 71.27.+a}
\vskip2pc

\maketitle

\section{Introduction}
Magnetic ions in Mott insulators often interact with short-ranged
antiferromagnetic exchange couplings $J$. When ions with spin $S$ occupy a bipartite
lattice, the system develops antiferromagnetic order with
the spins forming a collinear N\'eel state below a transition
temperature $T_N$  of order the Curie-Weiss temperature $\Theta_{CW} \sim JS^2$. However, 
if magnetic lattice defined by the nearest
neighbour connectivity matrix of the magnetic ions has
{\em triangular} loops (more generally, loops with an odd number
of sites), the antiferromagnetism is {\em frustrated}, and the system
does not develop long-range N\'eel order. 

In such cases, the classical exchange energy typically possesses
a very large number of minima---indeed, the ensemble defined by
the classical minimum-energy configurations often has finite entropy
in the thermodynamic limit. This leads to ``spin-liquid'' behaviour over a broad temperature range $T_f \ll T  \ll \Theta_{CW}$ in which the physical
properties reflect averages over all possible minimum energy configurations,
and are therefore largely determined by the geometry of the magnetic lattice.
Below the freezing temperature $T_f$, the system eventually orders, but
the ordering patterns are often quite complex and 
determined by the interplay between subleading terms in the interaction Hamiltonian and the effects of quantum and thermal fluctuations. \cite{Moessner_Ramirez}

In the spin-liquid regime, the geometric frustration effectively
``quenches''the leading exchange interactions---in some ways,
this is similar to fractional quantum hall systems in which
the formation of Landau levels quenches the leading kinetic energy
term in the Hamiltonian. As in the fractional quantum hall case,
this can lead to unusual, emergent degrees of freedom dominating
the physical response of the system.\cite{Rajaraman_fractionalization} One well-studied example
of this is the low temperature physics of spin-ice compounds in the
classical regime, which admits a natural description in terms of emergent magnetic monopole degrees of freedom in a classical easy-axis magnet.\cite{Castelnovo_Moessner_Sondhi} Another {\em quantum-mechanical} realization of such {\em quasiparticle
fractionalization} is the Coulomb liquid phase,\cite{Hermele_Balents_Fisher,Banerjee_etal,Shannon_etal,HusKrauMoesSond} e.g. of bosonic
matter on the three-dimensional pyrochlore lattice.

Here, we focus on the physics of the frustrated antiferromagnet
 SrCr$_{9p}$Ga$_{12-9p}$O$_{19}$ (SCGO),\cite{ober1,ober2,ramirez,kerenMUSR,schifferSUS,Limot_etal_prb} in which the Cr$^{3+}$ $S=3/2$
ions  interact with nearest neighbour Heisenberg
exchange interations $J \approx 80K$ to form a corner-sharing network of
antiferromagnetically coupled spins
that can be variously described as a Kagome bilayer or a pyrochlore
slab lattice whose sites are diluted with a density $x=1-p$ of
vacancies; these vacancies reflect the presence of non-magnetic Ga ions on the Cr sites in SCGO for $p <1$. Using the classical Heisenberg antiferromagnet on this SCGO lattice
to model the low temperature physics, we demonstrate that
the experimentally hitherto unachievable pure compound with $p=1$ remains
a spin-liquid down to the lowest temperatures accessible in classical
Monte-Carlo (MC) simulations; this is in keeping with analytical expectations.\cite{chalker1} 

We also demonstrate that the low temperature magnetic response of the site-diluted
system with dilution $x$ is most naturally described
in terms of the properties of spatially extended spin textures that 
cloak ``orphan'' $S=3/2$ Cr$^{3+}$ spins in direct proximity to a pair of
missing sites belonging to the same triangular simplex. 
Our
Monte-Carlo studies establish that a single orphan-texture complex in the low temperature limit has a net moment equal to exactly
half the moment of a free $S=3/2$ spin. Roughly speaking, this arises from
the fact that the orphan spin ``sees'' an effective magnetic field $h/2$ upon
application of a uniform external field $h$---half the external field being ``screened'' by the exchange-coupling
of the orphan spin to the surrounding spin-liquid---and the surrounding spin texture ``cancels off'' exactly half of
the orphan spin's magnetization by developing a net diamagnetic response
of the right magnitude in the low temperature limit.

This ``fractional moment'' is in agreement with predictions
of an effective field theory approach that we develop here. Furthermore, as already noted in
our earlier Letter\cite{Sen_Damle_Moessner_PRL}, the asymptotic low temperature
behaviour predicted by this effective theory is found to be surprisingly robust in MC simulations, persisting
up to temperatures of order $0.1JS^2$ for spin-$S$ magnets on the SCGO lattice with
nearest neighbour coupling $J$. Here, we show that the underlying reason
for this robustness has to do with the fact that this fractional moment of an individual
orphan-texture complex at $T=0$ is, in a well-defined sense, {\em much more localized} than
the $1/|\vec{r}|$ far-field fall-off of the surrounding spin texture.

At finite density, these textures (here and henceforth, we slur over the distinction
between the orphan-texture complex and the texture unless absolutely essential) interact with each other by an emergent {\em temperature dependent} {\em pair-wise exchange interaction} $J_{eff}(T,\vec{r})$ that decays exponentially beyond a thermal correlation length $\xi(T) \sim 1/\sqrt{T}$ and whose sign depends on whether the two orphan spins belong to the same
Kagome layer or not. Again, the absence of three-body and higher
interactions, as well as the dependence of $J_{eff}$ on $T$, $\vec{r}$
and layer index seen in MC simulations are all successfully modeled using the effective
field theory approach we develop here.  

Thus the physics of such vacancy-pairs {\em dominates} the impurity contribution to the
susceptibility of the Kagome bilayer at small $x$ in the low temperature limit. Although the contribution of these orphan spin textures is masked in macroscopic susceptibility measurements by another
``extrinsic'' contribution (see Sec.~\ref{sec:historySCGO}) that has little
to do with the physics of the frustrated Kagome bilayer, these spin textures
play a crucial role in determining Knight shifts and lineshapes in Ga NMR, as has already been emphasized in our previous Letter.\cite{Sen_Damle_Moessner_PRL}

This article is organized as follows. Sec.~\ref{sec:historySCGO} provides a whistle-stop tour of the most pertinent previous theoretical work. Our main technical tools, including the effective field theory approach we develop here, are described in some detail in Sec.~\ref{sec:methodsSCGO} and Sec.~\ref{fieldtheorycomputations}, which may be skipped
by a reader only interested in final results of relevance to SCGO. The following sections~\ref{sec:pure}, \ref{impurityeffects} contain a description of such results, first for the pure system, then for a single texture, and finally for interactions between the textures. The article closes with considerations on the generality of our results---in particular, how to extend them to other lattice topologies and dimensionalities---along with an outlook on 
future work.

\section{Background: Classical Heisenberg Model on the SCGO lattice}
\label{sec:historySCGO}
\begin{figure}
{\includegraphics[width=\hsize]{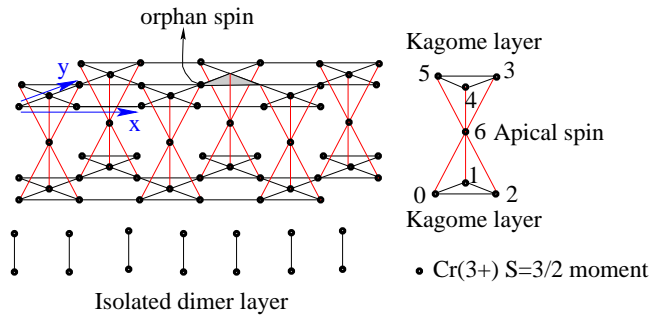}}
\caption{The Cr atoms in
SCGO form a lattice of Kagome bilayers separated from each other
by a layer of isolated dimers consisting of pairs of Cr atoms. Each Kagome
bilayer is a corner sharing arrangement of tetrahedra and triangles made up 
of two Kagome lattices coupled to each other through ``apical'' 
sites shared between up-pointing and down-pointing tetrahedra.
Links between near-neighbour sites in each bilayer represent a Heisenberg
exchange coupling $J = 80$ K between neighbouring Cr$^{3+}$ spins, while
links in the isolated dimer layer represent a Heisenberg exchange coupling
$J' = 216$ K between the two Cr$^{3+}$ ions that constitute each pair. Two
vacancies in the (shaded) triangle leave behind an orphan spin.}
\label{Fig1_lattice}
\end{figure}

We consider classical length-$S$ spins interacting with nearest neighbor Heisenberg exchange couplings on the SCGO lattice (see Fig \ref{Fig1_lattice}):

\be 
H_{nn} = J\sum_{\langle \vec{r} \vec{r}^{'} \rangle} \vec{S}_{\vec{r}} \cdot \vec{S}_{\vec{r}^{'}} - \sum_{\vec{r}} \vec{h}\cdot \vec{S}_{\vec{r}}
\label{eq1}
\ee
Here $J > 0$ is the antiferromagnetic Heisenberg exchange interaction on
a link $\langle \vec{r} \vec{r}^{'} \rangle$ connecting nearest neighbour sites
$\vec{r}$ and $\vec{r}^{'}$ on the SCGO lattice, and $\vec{h}$ is the uniform external magnetic field.

As depicted in Fig.~\ref{Fig1_lattice}, the Cr$^{3+}$ ions in SCGO define a  lattice consisting
of a pyrochlore slab with one layer
of up-pointing tetrahedra sharing a vertex each with a second layer of
down-pointing tetrahedra. Thus, this Kagome bilayer or pyrochlore
slab may be thought of as
a corner sharing network of up-pointing tetrahedra, down-pointing
tetrahedra, and triangles that do not belong to any tetrahedron---the
latter are the would-be bases of the next layers of tetrahedra
which would be needed to convert the SCGO lattice into a three dimensional
pyrochlore structure. These next layers of tetrahedra are absent in
the SCGO structure, being replaced in
effect by a layer of magnetically ``inert'' Cr dimers (spin pairs)
that serve to isolate each Kagome bilayer from the Kagome bilayer
in the next unit cell, and lead to an effectively two dimensional
situation. The main effect of this layer of Cr dimers is to give rise to a population of isolated Cr spin $S=3/2$ moments with density proportional to $x$ when
Ga impurities replace an  ${\mathcal O}(x)$ fraction of Cr spins in
this isolated dimer layer. Although the resulting population of free
Cr spins gives a dominant contribution to the macroscopic susceptibility
at small $x$ and low temperature, this has nothing to do with the physics of the frustrated Kagome bilayer. 
Therefore, we leave the Cr dimer layer out
of our discussion in the rest of this article.

Keeping this in mind, we rewrite the classical Hamiltonian
of the system as
\be
 {\mathcal H}(\{\vec{S}\}) = \frac{J}{2}\sum_{\XBox} (\sum_{\vec{r} \in \XBox}\vec{S}_{\vec{r}} - \frac{\vec{h}}{2J})^2 + \frac{J}{2}\sum_{\triangle} (\sum_{\vec{r} \in \triangle}\vec{S}_{\vec{r}} - \frac{\vec{h}}{2J})^2 \nonumber \\ 
\label{eq2}
\ee
where $\XBox$ and $\triangle$ denote the basic tetrahedral and triangular
simplical units of the kagome bilayers in the SCGO lattice (Fig \ref{Fig1_lattice}). It is now clear that
this classical exchange energy has a global minimum when each simplex of the kagome bilayer satisfies
\begin{eqnarray}
S^{\alpha}_{\XBox} \equiv \sum_{\vec{r} \in \XBox} S_{\vec{r}}^{\alpha} &=& \frac{h^{\alpha}}{2J} \mbox{~~~} \mathrm{and} \nonumber \nonumber \\
S^{\alpha}_{\triangle} \equiv \sum_{\vec{r} \in \triangle} S_{\vec{r}}^{\alpha} = \frac{h^{\alpha}}{2J} 
\label{eq3}
\end{eqnarray}
for each spin component $\alpha$. At zero magnetic field,
this implies that the total spin of each corner-sharing simplex is zero.

The $T=0$, $h=0$ physics is dominated by the ensemble
of states which achieve this global minimization of the energy by setting
the total spin of each simplex to zero, and this is
expected\cite{Moessner_berlinsky,Henley_2000,Henley_effectivetheory} to be the case even
in the presence of site dilution: this ansatz of ``simplex satisfaction'', whereby the total spin
of each simplex is set to zero in the configurations that
dominate the $T=0$ physics, has dramatic consequences. If any
simplex has more than one spin remaining on it after dilution, the
corresponding simplex continues to have zero total spin, while
a ``defective simplex'' with only one surviving spin on it {\em cannot}
have zero total spin. As a result, an infinitesimal magnetic
field acting at $T=0$ in the $\hat{z}$ direction only couples to such defective simplices, yielding the following saturation magnetization for
a system with a {\em single} defective simplex: $S^z_{tot} = \frac{1}{2}(\sum_{\XBox} S^z_{\XBox} + \sum_{\triangle} S^{z}_{\triangle}) = S/2$.

Thus, a single defective simplex leads to a magnetization 
$S/2$ in response to an infinitesimal magnetic field at $T=0$, hinting
at the presence of {\em fractionalized} spin degrees of freedom liberated
by such vacancy configurations.
Although this argument is explicitly restricted to $T=0$, and although
it is clear that spin-correlations at low temperature extend over
large distances in SCGO\cite{Henley_2000,Henley_effectivetheory}, it is curious to note that precisely these defective simplices give rise to a low temperature Curie response
$\propto 1/T$ within the ``single-unit approximation''
of Moessner and Berlinsky that does not explicitly incorporate spin-correlations beyond a single simplex.

Is this a coincidence, or are fractionalized spin $S/2$ degrees of
freedom ``real'' in the low temperature classical spin-liquid
regime in SCGO and related systems? Here, we address this question
using classical Monte-Carlo simulations of the microscopic system
as well as an effective field theory approach that correctly incorporates
entropic effects on the same footing as the energetics of simplex satisfaction. We will 
see that this approach enables us to describe the physics of vacancies with a 
remarkable level of accuracy and analytical detail.

\section{Methods: Effective field theory and classical Monte-Carlo}
\label{sec:methodsSCGO}
Here, we provide an overview of the effective field theory
and classical Monte-Carlo methods we use to explore the physics
of orphan spins in SCGO. The reader not interested in technical details may skip this section.

\subsection{Monte Carlo simulations}
In order to test our theoretical
predictions, we employ detailed classical Monte Carlo simulations to
study Eq.~\ref{eq1} with $O(3)$ and $O(4)$ spins. To update a spin configuration
efficiently, we use\cite{Lee_Young} a combination of ``microcanonical'' moves that conserve energy and the usual ``canonical'' Monte-Carlo moves.
In all cases, we use periodic
boundary conditions, and study a sequence of sizes ranging from $L=16$ to $L=50$ (with $7L^2$ sites)
to control finite-size effects and reliably extrapolate to the large
$L$ limit. In runs that focus on autocorrelation properties, we
are careful to ``switch off'' the microcanonical moves to guard against
the possibility that they mask glassy slowing down.

\subsection{Field theory: large-$N$ and soft spins}

In the pure case, 
the basic idea of the effective field theory is to treat the fixed length
constraint $\vec{S}_{\vec{r}}^2 = S^2$ in an approximate way by replacing the
original fixed length spins by fields $\vec{\phi}_{\vec{r}}$ whose {\em average}
length is controlled by a phenomenological stiffness parameter
to ensure that $\langle \vec{\phi}_{\vec{r}}^2\rangle = S^2$.
Thus, the partition function $Z_{{\mathrm eff}}$ of the effective theory for the pure
system is given by
\begin{eqnarray}
Z_{\mathrm{eff}} &=& \int {\mathcal D}\vec{\phi} \exp(-{\mathcal S}_{\mathrm{eff}}) \nonumber \\
\mathcal{S}_{\mathrm{eff}}&=&\frac{1}{2}\sum_{\vec{r}} \rho_{\vec{r}} \vec{\phi}_{\vec{r}}^2 + \frac{1}{T}{{\mathcal H}}\left(\{\vec{\phi}_{\vec{r}}\}\right) \nonumber \\
\label{eq4}
\end{eqnarray}
where ${\mathcal D} \vec{\phi} \equiv \prod_{\vec{r},\alpha} d\phi_{\vec{r}}^{\alpha}$, 
and the stiffness constants are fixed as $\rho_{\vec{r}}=\rho_1$ ($\rho_{\vec{r}}=\rho_2$)  for all the sites on the kagome (apical) layers by requiring that
$\langle \vec{\phi}^2_{\vec{r}}\rangle_{Z_{\mathrm{eff}}} = S^2$. 
Solving these constraints in the thermodynamic limit, we obtain  $\rho_1\approx 1.2350 S^2$ and $\rho_2 \approx 1.5903 S^2$ at $T=0$ and $h=0$. In the spin liquid regime ($T \ll JS^2,h \ll JS$), the changes to $\rho_1$ and $\rho_2$ induced by non-zero temperatures
and fields are small and can be ignored to a good approximation \cite{Garanin_Canals}. 

Although our vector field $\vec{\phi}_{\vec{r}}$ is a three component object defined
at each lattice site, this effective field theory treatment of the pure system is equivalent to the 
leading term in a large-$N$ expansion for the $O(N)$ generalization of the original classical
Heisenberg spin system. Our rationale for relying
on this approach is that this leading term in the large-$N$ expansion is
known to provide a
very good approximation to the physics of the classical Heisenberg
antiferromagnet on both the Kagom\'e and pyrochlore lattices in their
respective spin liquid regimes.\cite{Garanin_Canals,Isakov_Moessner_Sondhi}
Since the
SCGO lattice is essentially a finite-thickness slab of the pyrochlore
lattice, we expect that this large-$N$ treatment will be a good
approximation in the spin liquid regime on the SCGO lattice as well.
In what follows, we explicitly test this assumption against
the results of classical Monte Carlo simulations before proceeding
to study the effect of vacancies.

The fact that SCGO is a spin-liquid in this approximation can be seen 
by taking the $T \rightarrow 0$ of the action in Eq ~\ref{eq4}. Then, we get 
\be 
Z_{eff} = \int \mathcal{D}\vec{\phi}_{\vec{r}} \exp\left(-\frac{1}{2} \sum_{\vec{r}} \rho_{\vec{r}} \vec{\phi}_{\vec{r}}^2 \right) \prod_{\triangle_{\vec{r}}} \delta(\vec{\phi}_{\triangle_{\vec{r}}})\prod_{\XBox_{\vec{r}}} \delta(\vec{\phi}_{\XBox_{\vec{r}}}) \nonumber \\
\label{eq5}
\ee 
The local constraints on the ground state manifold ($\vec{\phi}_{\triangle_{\vec{r}}} \equiv \sum_{\vec{r} \in \triangle} \vec{\phi}_{\vec{r}} =0$, and
$\vec{\phi}_{\XBox_{\vec{r}}}\equiv \sum_{\vec{r} \in \XBox} \vec{\phi}_{\vec{r}}=0$) can be
elegantly described by constructing vector fields $\vec{B}^\alpha$ that live on the bonds of the dual lattice of the
SCGO lattice; the sites of this dual lattice are located at the centers of the simplices of the SCGO lattice,
and the links of this dual lattice pass through the SCGO sites. Thus we write
\be
\vec{B}^\alpha(\vec{r}) = \hat{e}_{\vec{r}}\phi^\alpha(\vec{r}), \mbox{~~~~~~~}\alpha=x,y,z
\label{eq6}
\ee
where the unit vectors $\hat{e}$ are always directed from the $A$ sublattice site of the (bipartite) dual lattice to
the neighbouring $B$ sublattice site. With this choice, these constraints translate to the requirement that each of the $\vec{B}^\alpha$ fields are divergence-free:
\be
\vec{\nabla} \cdot \vec{B}^\alpha=0
\ee
 and the action in Eq ~\ref{eq5} becomes 
\be
\frac{1}{2}\sum_{\vec{r}} \rho_{\vec{r}}(\vec{B}^{\alpha}_{\vec{r}})^2 
\ee
This implies spin correlators which are dipolar and decay as $1/r^2$ in real space. The zero-temperature formulation and behaviour of such a Coulomb phase is well understood in related settings. \cite{Henley_2000,Henley_effectivetheory,Isakov_Moessner_Sondhi}
At non-zero but small temperatures, there is a finite thermal correlation length $\xi \sim 1/\sqrt{T}$ beyond which the spin correlations decay exponentially. In Coulomb language, $\vec{\phi}_\triangle$ and $\vec{\phi}_{\XBox}$ can be thought of as ``vector-charges'', and these are strictly zero at $T=0$ but can be generated thermally, without an excitation gap, when $T \neq 0$. As we will see, the spin-``charge'' correlator of the undiluted parent spin liquid determines the texture induced around an orphan spin and the ``charge''-``charge'' correlator determines the interaction between orphan spins within our effective theory.

\subsection{Field theory: Vacancy effects}
\label{vacancies}
To handle the presence of vacancies, we employ Lagrange
multipliers $\lambda^{\alpha}_{\vec{r}}$, {\it i.e.} the Fourier representation 
\be 
\delta (\phi^\alpha_{\vec{r}}) = \frac{1}{2\pi}\int d\lambda_{\vec{r}}^\alpha \exp(i\lambda_{\vec{r}}^{\alpha} \phi_{\vec{r}}^\alpha)
\ee
to enforce the constraint $\phi_{\vec{r}}^{\alpha} = 0$
for each component $\alpha$ of $\vec{\phi}_{\vec{r}}$ at all sites $\vec{r}=\vec{r}_v$ on
which the Cr$^{3+}$ spins have been substituted by non-magnetic Ga. 
Beyond this, our central technical innovation lies in a careful treatment of those 
{\em defective simplices}, in which {\em all spins but one have been
substituted by non-magnetic Ga}.  It is abundantly clear 
that the fixed length constraint
on the microscopic spins is  crucial for the single ``orphan'' spin
left behind at the neighbouring site $\vec{r}_{o}$ on such defective simplices. In order correctly to capture
the physics of these orphan spins, we retain
the original microscopic fixed length spin-$S$ variable $\vec{S}_{\vec{r}_{o}}$ 
at each such orphan sites $\vec{r}_{o}$, {\it i.e.} we do {\em not} approximate
these orphan spins by Gaussian fields $\vec{\phi}_{\vec{r}_{o}}$.

Thinking in terms of the large-$N$ limit of the $O(N)$ spin model with vacancies
provides an alternate perspective on our approach: In the presence of
vacancies, there is no translational invariance, and the stiffness parameters
$\rho_{\vec{r}}$ of the large-$N$ theory acquire a dependence on the spatial
position $\vec{r}$. Since the most dramatic
spatial dependence of $\rho_{\vec{r}}$ is expected to arise at and perhaps around each orphan site $\vec{r}=\vec{r}_{o}$,  retaining these orphan spins as microscopic
fixed-length vectors is a way of incorporating
the leading effects of this spatial inhomogeneity. Having done this,
we expect that the stiffness $\rho_{\vec{r}}$ does not
deviate significantly from its uniform system value at {\em other} sites.
With this in mind, we ignore the $\vec{r}$ dependence of $\rho_{\vec{r}}$ once
the orphan spins are retained as spin-$S$ vectors. The validity of this assumptions
is confirmed {\em a posteriori} against numerics. 

Our effective theory thus reduces to a field theory
with a constrained Gaussian field $\phi_{\vec{r}}$ 
coupled to fixed-length spin-$S$ vectors at the orphan sites.
We are thus led to a sort of
``classical Kondo lattice model'' in which one has spin-$S$
variables coupled to a bosonic ``bath'' represented by $\vec{\phi}$.
It is interesting to note that the effective coupling
between the orphan spins, and the effective ``g-factor'' with which the
orphan spins couple to the external magnetic field,
are both determined by the coupling of orphan spins to the bosonic field $\vec{\phi}$,
which may be thought of as representing the degrees
of freedom of the bulk spin liquid.

Calculations within this framework are most conveniently performed
by noting that the fixed-length spin-$S$ character of the orphan
spins can also be implemented by introducing additional Lagrange multipliers
$\mu^{\alpha}_{\vec{r}_{o}}$ to enforce the constraint $\phi_{\vec{r}_{o}}^{\alpha} = Sn_{\vec{r}_{o}}^{\alpha}$, where $\vec{n}_{\vec{r}_{o}}$ is a unit vector at each orphan spin site $\vec{r}_{o}$ (so that $S\vec{n}_{\vec{r}_{o}}$ is the length-$S$ orphan spin vector). Note that these
Lagrange multipliers $\lambda$ and $\mu$ play a quite {\em different}
role in our theory from the stiffness parameters $\rho$: The $\rho$
capture the effects of the fixed length constraint {\em on average}
and are {\em not} integrated over, while the Lagrange multipliers $\lambda$
and $\mu$ are {\em integrated over in order to set $\vec{\phi}$ to $0$ at
vacancy sites and $S\vec{n}$ at the orphan sites}.
 
To avoid cluttering notation with factors of $J$ and $S$, we now measure temperature in units of $JS^2$
({\em i.e.} replace $T/JS^2$ by $T$) and magnetic field $h^{\alpha}$ in units of $JS$ ({\em i.e.} replace
$h^{\alpha}/JS$ by $h^{\alpha}$) and use $\langle \vec{\phi}^2 \rangle =1$.
In other words, we write
\begin{eqnarray}
Z_{\mathrm{eff}} &=& \int {\mathcal D} \vec{n}\int {\mathcal D}\vec{\lambda} \int {\mathcal D}\vec{\mu} \int {\mathcal D}\vec{\phi} \exp(-{\mathcal S}_{\mathrm{eff}}^{'}) \nonumber \\
\mathcal{S}_{\mathrm{eff}}^{'}&=& \mathcal{S}_{\mathrm{eff}} - i\sum_{\vec{r}_{v}\alpha} \lambda_{\vec{r}_{v}}^{\alpha} \phi_{\vec{r}_{v}}^{\alpha} - i \sum_{\vec{r}_{o} \alpha} \mu_{\vec{r}_{o}}^{\alpha} (\phi_{\vec{r}_{o}}^{\alpha} - n_{\vec{r}_{o}}^{\alpha}) \nonumber \\
\label{eq7}
\end{eqnarray}
where ${\mathcal D}\vec{\lambda} \equiv \prod_{\vec{r}_{v},\alpha} d\lambda_{\vec{r}_{v}}^{\alpha}$,
${\mathcal D}\vec{\mu} \equiv \prod_{\vec{r}_{o},\alpha} d\mu_{\vec{r}_{o}}^{\alpha}$,
and ${\mathcal D}\vec{n} \equiv \prod_{\vec{r}_{o},\alpha} dn_{\vec{r}_{o}}^{\alpha}\delta(\vec{n}_{\vec{r}_{o}}^2 -1)$.
Since all these Lagrange
multipliers $\lambda_{\vec{r}_{v}}$ and $\mu_{\vec{r}_{o}}$ couple linearly to the Gaussian field $\vec{\phi}$, and since ${\mathcal S}_{\mathrm{eff}}(\{\vec{\phi}_{\vec{r}}\},\vec{h},T)$ is the action
for the translationally invariant pure system, one
can do the Gaussian
integrals over the $\vec{\phi}_{\vec{r}}$ fields exactly by performing
the corresponding integrals over the Fourier modes of the
$\vec{\phi}$ field.

Upon doing these integrals, one obtains an action written
as a quadratic
form in terms of the uniform external magnetic field $h$ and the Lagrange
multiplier fields $\lambda^{\alpha}_{\vec{r}_{v}}$ and $\mu^{\alpha}_{\vec{r}_{o}}$ at vacancy sites
$\vec{r}_{v}$ and orphan sites $\vec{r}_{o}$ respectively. This quadratic form is of course diagonal
in the spin index $\alpha$, and $\vec{h}$ only appears in the action
for the $z$ component ($\alpha=z$) if the field acts along the $\hat{z}$ axis. Moreover, the fixed length orphan spins $n^{\alpha}_{\vec{r}_{o}}$ 
couple linearly to the $\mu_{\vec{r}_{o}}^{\alpha}$ and thus appear as
{\em sources} in this effective action. Thus, we obtain
\begin{widetext}
\begin{eqnarray}
Z_{\mathrm{eff}} & \propto & \int {\mathcal D} \vec{n}\int {\mathcal D}\vec{\lambda} \int {\mathcal D}\vec{\mu}
\exp \left(+\frac{1}{2}\sum_{\vec{r} \vec{r}^{'} \alpha}(\beta h^{\alpha} + i\Lambda_{\vec{r}}^{\alpha})(M^{-1})_{\vec{r} \vec{r}^{'}}(\beta h^{\alpha} + i\Lambda_{\vec{r}^{'}}^{\alpha}) - i \sum_{\vec{r}_{o} \alpha} \mu_{\vec{r}_{o}}^{\alpha} n_{\vec{r}_{o}}^{\alpha}\right)
\label{eq8}
\end{eqnarray}
\end{widetext}
Here the matrix $M$ is defined by rewriting the quadratic part of ${\mathcal S}_{\mathrm{eff}}$ as 
\be 
\sum_{\vec{r} \vec{r}^{'} \alpha}\phi_{\vec{r}}^{\alpha} M_{\vec{r} \vec{r}^{'}} \phi_{\vec{r}^{'}}^{\alpha}/2 \; ,
\ee 
 $\beta$ is the inverse temperature,
\be 
\Lambda_{\vec{r}}^{\alpha} =  \delta_{\vec{r}, \vec{r}_{v}}\lambda_{\vec{r}_v}^{\alpha} +  \delta_{\vec{r}, \vec{r}_{o}}\mu_{\vec{r}_{o}}^{\alpha}\; ,
\ee 
 and $h^{\alpha}$ may be assumed to be of the form
$h^{\alpha} = h \delta_{\alpha z}$ without loss of generality.

Crucially, this may be rewritten {\em in terms of the susceptibility matrix ${\mathbf \chi}$
whose elements $\chi_{\vec{r} \vec{r}^{'}}(T)$ are the effective field theory prediction
for the linear susceptibility 
of the spin at site $\vec{r}$ of the pure system in response to a local field applied
at site $\vec{r}^{'}$}. To see this, we first note
that $M^{-1}$ can be identified with
\begin{eqnarray}
\langle \phi_{\vec{r}}^{\alpha} \phi_{\vec{r}^{'}}^{\beta} \rangle & \equiv & C_{\vec{r} \vec{r}^{'}} \delta_{\alpha \beta} \; ,
\label{eq9}
\end{eqnarray}
{\em the zero field correlations of the $\vec{\phi}$ field in the pure problem}:
\begin{eqnarray}
 (M^{-1})_{\vec{r} \vec{r}^{'}} &= & C_{\vec{r} \vec{r}^{'}} 
\label{eq10}
\end{eqnarray}
Furthermore, we have quite generally
\begin{eqnarray}
\chi_{\vec{r} \vec{r}^{'}} & = & \beta C_{\vec{r} \vec{r}^{'}}
\label{eq11}
\end{eqnarray}
and therefore, we may write
\begin{eqnarray}
 (M^{-1})_{\vec{r} \vec{r}^{'}} &= & T \chi_{\vec{r} \vec{r}^{'}} 
\label{eq12}
\end{eqnarray}

In other words, our effective action can be rewritten as
\begin{widetext}
\begin{eqnarray}
Z_{\mathrm{eff}} & \propto & \int {\mathcal D} \vec{n}\int {\mathcal D}\vec{\lambda} \int {\mathcal D}\vec{\mu}
\exp \left(+\frac{\beta}{2}\sum_{\vec{r} \vec{r}^{'} \alpha}(h^{\alpha} + iT\Lambda_{\vec{r}}^{\alpha})\chi_{\vec{r} \vec{r}^{'}}(h^{\alpha} + iT\Lambda_{\vec{r}^{'}}^{\alpha}) - i \sum_{\vec{r}_{o} \alpha} \mu_{\vec{r}_{o}}^{\alpha} n_{\vec{r}_{o}}^{\alpha}\right) \nonumber \\
&=& \int {\mathcal D} \vec{n}\int {\mathcal D}\vec{\lambda} \int {\mathcal D}\vec{\mu}
\exp \left(\frac{\beta\vec{h}^2}{2}\sum_{\vec{r}\vec{r}^{'}} \chi_{\vec{r}\vec{r}^{'}}-\frac{1}{2}\sum_{\vec{r} \vec{r}^{'} \alpha}\Lambda_{\vec{r}}^{\alpha}C_{\vec{r} \vec{r}^{'}}\Lambda_{\vec{r}^{'}}^{\alpha} +i\sum_{\vec{r} \alpha}\Lambda_{\vec{r}}^{\alpha}(h^{\alpha}(\sum_{\vec{r}^{'}}\chi_{\vec{r} \vec{r}^{'}})- \delta_{\vec{r} \vec{r}_{o}}n_{\vec{r}_{o}}^{\alpha} )\right)
\label{eq13} 
\end{eqnarray}
\end{widetext}
where we have written the quadratic part of the action for $\Lambda$ in
terms of the correlation matrix $C$, while writing other terms using
the susceptibility matrix $\chi$, since this proves to
be the most physically transparent representation for the
subsequent analysis.

One may now integrate over the Lagrange multipliers $\lambda^{\alpha}_{\vec{r}_{v}}$ and
$\mu^{\alpha}_{\vec{r}_{o}}$ to obtain an effective action
that couples all the orphan spins $Sn^{\alpha}_{\vec{r}_{o}}$ to each other and to the external magnetic field $h$. The structure of the resulting
effective action for the $\vec{n}$ is controlled by
the structure of the quadratic form ${\mathcal P}_{\Lambda}C {\mathcal P}_{\Lambda}$, where the projection
operator ${\mathcal P}_{\Lambda}$ restricts $C$ to the
subspace spanned by $\vec{r}$ for which $\Lambda_{\vec{r}}^{\alpha}$ is non-zero, {\it i.e}
$\vec{r}$ corresponding to orphan sites $\vec{r}_{o}$ and vacancy sites $\vec{r}_{v}$.  

From knowledge of this quadratic form, which depends crucially on the disorder configuration, 
one can compute $\langle n^z_{\vec{r}_{o}} \rangle (h,T)$ of each orphan spin at a given temperature and field. 
Finally, by
analyzing the path integral over $\vec{\phi}$ {\em with the orphan spins 
replaced by sources of strength equal to the orphan spin polarization
$\langle n^{z}_{\vec{r}_{o}}\rangle(h,T)$}, we obtain $\langle \phi^{z}_{\vec{r}}\rangle(h,T)$, the spin texture induced around itself by each orphan.

In Section~\ref{impurityeffects}, we summarize the results
of such an analysis for one, two and three orphan configurations,
relegating details to Section~\ref{fieldtheorycomputations}
and Appendix~\ref{fieldtheoryappendix}.

\section{Calculations within effective field theory}
\label{fieldtheorycomputations}

Since the effective field theory approach outlined above is somewhat novel,
it is perhaps useful to first address some ``structural'' questions regarding the overall logic of this approach and its
consistency before we turn to the actual calculations.

\subsection{Treating one spin as a unit vector in absence of orphans}
To this end, we first consider the pure system and ask:
What would this approach predict if we chose to retain any one spin, say the one
at site $\vec{r}$, as a fixed-length vector $\vec{n}$, while
using the effective field theory description for the rest
of the system? To answer this, one simply notes that
$C_{\vec{r}r} = 1/3$ independent of $T$, and that $\sum_{\vec{r}^{'}} \chi_{\vec{r} \vec{r}^{'}} = \chi_{\vec{r}}(T)$, the
susceptibility per site obtained within the effective field theory. $\chi_{\vec{r}}(T)$ equals $\chi_{kag}(T)$ (susceptibility of a site in the kagome layers) or $\chi_{ap}(T)$ (susceptibility of a site in the apical layer) depending on the location of $\vec{r}$.
Our approach would give
\begin{widetext}
\begin{eqnarray}
Z_{\mathrm{eff}} &\propto& \int d\vec{n} \delta(\vec{n}^2-1)\int d\vec{\mu}
\exp \left(\frac{\beta\vec{h}^2}{2}\sum_{\vec{r}\vec{r}^{'}} \chi_{\vec{r}\vec{r}^{'}}-\frac{1}{6}\sum_{\alpha}\vec{\mu}^2 +i\vec{\mu}\cdot(\chi_{\vec{r}}\vec{h}- \vec{n}) \right)  \nonumber \\
&\propto& \int d\vec{n} \delta(\vec{n}^2-1) \exp (3  \chi_{\vec{r}} \vec{h} \cdot \vec{n})
\label{eq14}
\end{eqnarray}
\end{widetext}
Note that the ${\mathcal O}(\vec{n}^2)$ terms add up to a constant because of the unit length constraint on $\vec{n}$. From the above equation, we get that $\langle n^z \rangle = \left(\frac{-1}{3 \chi_{\vec{r}} h}+\coth(3 \chi_{\vec{r}} h)\right)$ which reduces to $\langle  n^z \rangle = \chi_{\vec{r}} h$ in the low-field limit.

Another natural question that comes to mind is the following: What
does this approach predict for an arbitrary site $\vec{r}$ if we
choose to retain the spin at that site as a fixed length vector
$\vec{n}$ in a sample with a {\em single vacancy} at the origin?
${\mathcal P}_{\Lambda} C {\mathcal P}_{\Lambda}$ is now
two dimensional (corresponding to the vacancy site $0$ and the site $\vec{r}$),
and we obtain the following effective action for the spin at $\vec{r}$:
\begin{equation}
Z_{\mathrm{eff}} \propto \int d\vec{n} \delta(\vec{n}^2-1) \exp ( C_{\mathrm{eff}}(\vec{r},0)\vec{h} \cdot \vec{n})
\label{eq15}
\end{equation}

where $C_{\mathrm{eff}}(\vec{r},0) = (\chi_{\vec{r}} -3C_{0\vec{r}}\chi_0)/(\frac{1}{3}-3C_{0\vec{r}}^2)$. Crucially $C_{\mathrm{eff}}(\vec{r},0)$ stays finite for all $\vec{r}$ as $T \rightarrow 0$ ruling out Curie response of any kind in the single vacancy case. When $\vec{r}$ is well seperated from $0$, we can ignore $C_{0\vec{r}}$ and $C_{\mathrm{eff}}=3\chi_{\vec{r}}$, which is identical to the result for the undiluted problem derived above.  When both sites $\vec{r}$ and $0$ belong to one of the kagome layers, $C_{\mathrm{eff}}$ gets further simplied to $C_{\mathrm{eff}}=\frac{3 \chi_{kag}}{1+3C_{0\vec{r}}}$. Thus the local magnetization $\langle  n^z(\vec{r}) \rangle$ at low $T$ and uniform field $h$ encodes the information about the spin-spin correlations of the parent spin liquid (i.e. $C_{0\vec{r}}$) in the case of a single vacancy. However, the orphan spins lead to {\em parametrically} stronger effects and the effects of single vacancies in a diluted lattice can be ignored to a good approximation.

In both these examples (no vacancy and a single vacancy), treating one spin as a fixed-length vector while
describing the rest of the system by a Gaussian field resulted
in a prediction that the spin that was singled out continues to
have a finite magnetic suscepbility in the low temperature limit.
When would our approach predict a Curie-like response with
a divergent susceptibility in the low temperature limit?

The answer has to do with the low temperature limit of the eigenspectrum of ${\mathcal P}_{\Lambda} C{\mathcal P}_{\Lambda}$: When the projector ${\mathcal P}_{\Lambda}$
projects onto the subspace spanned by all sites of a simplex (i.e three
sites that form a triangular simplex, or four sites that form
a tetrahedral simplex), it is easy to see that the resulting
${\mathcal P}_{\Lambda} C{\mathcal P}_{\Lambda}$ has one eigenvalue which
goes to zero with temperature. The corresponding
eigenvector is the uniform eigenvector with equal amplitude
at all sites of the simplex---this is in effect a consequence of
the fact that the Hamiltonian ${{\mathcal H}}\left(\{\vec{\phi}_{\vec{r}}\}\right)$ enforces the constraints
\be
\sum_{\vec{r} \in \XBox} \phi_{\vec{r}}^{\alpha} = 0 \mbox{~~~} \mathrm{and} \mbox{~~~} \sum_{\vec{r} \in \triangle} \phi_{\vec{r}}^{\alpha} = 0  
\label{eq16}
\ee
which in turn guarantees the fact that the vector with equal
amplitude at all sites of a simplex is a zero mode of ${\mathcal P}_{\Lambda} C{\mathcal P}_{\Lambda}$. At finite temperature $T$, 
the precise statement is that
the eigenvalue $\epsilon_0$ corresponding to this mode scales with
temperature as $\epsilon_0 = T$.

When there is more than one orphan spin present in the system, 
every defective simplex gives rise to a Curie-tail in the 
spin susceptibility even
in the presence of the other defective simplices---the correlations
between different defective simplices will show up as entropically
generated effective interactions between the orphan spins on
these defective simplices; these are the effective exchange
couplings $J_{eff}$ that we calculate later in this article.
Obtaining the low temperature behaviour of the orphan spins $S\vec{n}_{\vec{r}_{o}}$
from knowledge of these interactions, one can in principle feed this information back in
to compute the expected response of the surrounding spin liquid, and thus obtain
the low-temperature properties of {\em this system of interacting orphan-texture complexes}.

\subsection{A single-simplex embedded in the spin-liquid}
Another way to re-phrase the argument is to consider three unit-length vectors on the three sites of a triangular simplex. We can then integrate out the rest of the soft spin degrees of freedom to obtain an effective action for these three spins on the simplex. This is then the ``single-unit'' action in this effective field theory description:
\begin{widetext}
\begin{equation}
Z_{\mathrm{eff}} \propto \int d\vec{n_1} d\vec{n_2} d\vec{n_3} \exp \left( \frac{3\chi_{kag}}{\langle (\phi_\triangle^z)^2\rangle} \vec{h} \cdot (\vec{n}_1+\vec{n}_2+\vec{n}_3)+\frac{9 C_{nn}}{1-3C_{nn}}\frac{\vec{n}_1 \cdot \vec{n}_2 +\vec{n}_2 \cdot \vec{n}_3+\vec{n}_3 \cdot \vec{n}_1 }{\langle (\phi_\triangle^z)^2\rangle}\right)   
\label{eq17}
\end{equation}   
\end{widetext}

where both $C_{nn}$ (nearest neighbor correlation on the kagome layer) and $\langle (\phi_\triangle^z)^2\rangle$ are calculated in the parent spin liquid state at $h=0$. At low $T$, the above effective action can be simplied by noting that $\langle (\phi_\triangle^z)^2\rangle = T$ by equipartition, $C_{nn} = -\frac{1}{6}+O(T)$ and $\chi_{kag} = \frac{1}{6}+O(T)$.
Then, we get the following action
\begin{widetext}
\begin{equation}
Z_{\mathrm{eff}} \propto \int d\vec{n_1} d\vec{n_2} d\vec{n_3} \exp \left( \frac{1}{2} \beta \vec{h} \cdot (\vec{n}_1+\vec{n}_2+\vec{n}_3)- \beta (\vec{n}_1 \cdot \vec{n}_2 +\vec{n}_2 \cdot \vec{n}_3+\vec{n}_3 \cdot \vec{n}_1 )\right)  
\label{eq18} 
\end{equation}   
\end{widetext}
which immediately shows that removing two (but not one) spins from the triangular simplex leads to the absence of all the pair-wise interaction terms above and the surviving $\vec{h} \cdot \vec{n}$ term then leads to the Curie response of the orphan spin. Also note that at low temperatures, the orphan spin acts like a free spin in a magnetic field of ``$h/2$'' instead of the applied field $h$. This is a direct consequence of the rest of the field being screened by the coupling to the surrounding spin liquid.

\subsection{Texture induced by orphan spin}
The induced spin texture at a point $\vec{r}_2$ far away from an orphan spin located at $\vec{r}_1$ can be expressed simply in terms of the parent spin liquid properties: When $|\vec{r}_2-\vec{r}_1| \gg a$, where $a$ is the lattice spacing, the detailed ``internal'' structure of the orphan spin simplex becomes irrelevant and we can simply impose $\vec{\phi}_{\triangle}(\vec{r}_1)=\vec{n}_{\mathrm{orphan}}$ instead of removing two vacancies from the simplex. To calculate the spin polarization at $\vec{r}_2$, it is useful to first integrate out all other ($r \neq \vec{r}_2$) $\vec{\phi}(\vec{r})$ to obtain an effective action that couples $\vec{\phi}(\vec{r}_2)$ to the unit vector $\vec{n}_{\rm orphan}$. In practice, we do this by integrating over {\it all} $\vec{\phi}(\vec{r})$ while constraining $\vec{\phi}(\vec{r}_2)$ to take on a fixed value. This gives
\begin{widetext}
\be 
Z_{eff} \propto \int \mathcal{D} \vec{n}_{orphan} \mathcal{D}\vec{\phi}(\vec{r}_2) \mathcal{D}\vec{\mu}\exp \left(-\frac{1}{2}\sum_{\vec{r},\vec{r}^{'}=\vec{r}_1,\vec{r}_2} \mu_{\vec{r}}^\alpha \mathcal{M}^1_{\vec{r}\vec{r}^{'}}\mu_{\vec{r}^{'}}^\alpha + i(3\chi_{kag}h^{\alpha}-n^{\alpha}_{\rm orphan}(\vec{r}_1))\mu_{\vec{r}_1}^\alpha +i(\chi_{\vec{r}_2}h^{\alpha}-\phi^{\alpha}(\vec{r}_2))\mu_{\vec{r}_2}^\alpha \right) \nn \\
&&
\label{eq19}
\ee 
\end{widetext}
where $\mu_{\vec{r}}^\alpha$ at $r=\vec{r}_1,\vec{r}_2$ respectively impose the constraints that $\vec{\phi}_{\triangle}$ on the  ``orphan'' triangle at $\vec{r}_1$ equals $\vec{n}_{\rm orphan}$ and $\vec{\phi}(\vec{r}_2)$ is held fixed. In the above, the matrix 
\begin{displaymath}
\mathcal{M}^1= 
\left( \begin{array}{cc}
\frac{\langle \vec{\phi}_{\triangle}^2(\vec{r}_1) \rangle}{3} &\frac{ \langle\vec{\phi}_{\triangle}(\vec{r}_1)\cdot \vec{\phi}(\vec{r}_2) \rangle}{3}\\
\frac{ \langle\vec{\phi}_{\triangle}(\vec{r}_1)\cdot \vec{\phi}(\vec{r}_2) \rangle}{3} & \frac{1}{3}  
\end{array} \right)
\end{displaymath}
is our approximation to $P_{\Lambda}CP_{\Lambda}$ obtained by ignoring the internal structure of the orphan spin simplex
(correlators appearing in the matrix elements represent correlations of the pure spin effective field theory
in zero field). Performing the integrals over the $\vec{\mu}$ fields, we finally obtain
\begin{widetext}
\be
Z_{eff} & \propto & \int \mathcal{D} \vec{n}_{\rm orphan} (\vec{r}_1) \mathcal{D} \vec{\phi} (\vec{r}_2) \exp 
\left( 
\frac{3 \langle \vec{\phi}_{\triangle} (\vec{r}_1) \cdot \vec{\phi} (\vec{r}_2) \rangle}{\langle \vec{\phi}^2_{\triangle} \rangle - \langle \vec{\phi}_{\triangle} (\vec{r}_1) \cdot \vec{\phi} (\vec{r}_2) \rangle^2} \vec{n}_{\rm orphan} (\vec{r}_1) \cdot \vec{\phi}(\vec{r}_2) + \frac{9 \chi_{kag}}{\langle \vec{\phi}^2_{\triangle} \rangle} \vec{h} \cdot \vec{n}_{\rm orphan} (\vec{r}_1) 
\right. \\ \nn
& + & 
\left. 
3\chi(\vec{r}_2)\vec{h}\cdot \vec{\phi}(\vec{r}_2) -\frac{3\langle \vec{\phi}_{\triangle}^2 \rangle}{2(\langle \vec{\phi}^2_{\triangle}\rangle - \langle\vec{\phi}_{\triangle}(\vec{r}_1) \cdot \vec{\phi}(\vec{r}_2) \rangle^2)}\vec{\phi}(\vec{r}_2)\cdot \vec{\phi}(\vec{r}_2)
\right)
\label{eq20}
\ee
\end{widetext}

We now compute both $\langle n^z_{\rm orphan}(\vec{r}_1) \rangle$ and $\langle \phi^z(\vec{r}_2) \rangle$ from this effective action. At low temperature, we obtain $\langle n^z_{\rm orphan}(\vec{r}_1) \rangle = \mathcal{B}(h/2,T)$, the polarization
of a unit-length spin at temperature $T$ in response to an external field of magnitude $h/2$; this was also obtained from the more detailed calculation in the previous section where the two vacancies on a triangular simplex were explictly incorporated to compute the orphan spin response. At vanishingly small
fields $h$ deep in the low temperature spin-liquid regime ($h \ll T \ll 1 $),
we obtain:
\be
\langle \phi^z(\vec{r}_2) \rangle \approx \chi(\vec{r}_2)h + \frac{\langle \phi^z_{\triangle}(\vec{r}_1) \phi^z(\vec{r}_2) \rangle}{T}\mathcal{B}(h/2,T) ~~.
\label{eq21}
\ee

Thus, we see that the texture induced around an orphan spin is intimately related to the spin-``charge'' correlation function of the parent spin liquid (where $\vec{\phi}_\triangle$ is the thermally generated vector-``charge'' defined
earlier). In
the Appendix, we use a general long-wavelength analysis of the properties of a ``Coulomb-liquid'' with
fluctuating fields to gain insight into the nature of these spin-``charge'' correlations.

\subsection{Orphan spin interaction}
\label{orphans}
As the orphan spin textures are extended degrees of freedom, it is {\em a priori} not at all obvious how they will interact as the centers of two of them are brought closer together. The great advantage of the method we have developed above is that it readily generalizes to the case of any number of orphan spins, albeit with increased calculational effort. We document the details relevant to the two orphan case in Appendix.~\ref{app:orphint}.
The final form of the action is about as simple as could have been hoped for:
\begin{widetext}
\be 
\mathcal{Z} \propto \int \mathcal{D}\vec{n}_1 \mathcal{D}\vec{n}_2\exp \left(-\beta J_{eff}\vec{n}_1 \cdot \vec{n}_2+\beta  h s_1 n_{1}^{z}+\beta  h s_2 n_{2}^{z} \right) 
\label{eq22}
\ee
\end{widetext}
Basically, the two Zeeman terms for the individual orphan spins, where $s_1,s_2 \rightarrow 1/2$ at low $T$ are supplemented by an effective exchange `constant' $J_{eff}$, which depends on the location of the orphans, and on temperature. 

$\beta J_{eff}(\vec{r},T)$ can be appoximately expressed in an illuminating form, which becomes exact in the limit $r \gg a$, i.e. when the two orphan sites are well seperated. When $r \gg a$, we ignore the ``internal'' structure of the orphans as before and simply impose $\vec{\phi}_{\triangle}(\vec{r}_1)=\vec{n}_1$ and $\vec{\phi}_{\triangle}(\vec{r}_2)=\vec{n}_2$ instead of removing two spins from each of the triangular simplices on which the orphans reside. Then integrating out the remaining degrees of freedom yields the following effective action (here we stick to $h=0$ for notational simplicity):
\begin{widetext}
\be 
Z_{eff} \propto \int \mathcal{D} \vec{n} \mathcal{D}\vec{\mu} \delta(\vec{n}_1^2-1) \delta(\vec{n}_2^2-1)\exp \left(-\frac{1}{2}\sum_{\vec{r},\vec{r}^{'}=\vec{r}_1,\vec{r}_2} \mu_{\vec{r}}^\alpha \mathcal{M}^2_{\vec{r}\vec{r}^{'}}\mu_{\vec{r}^{'}}^\alpha + i\sum_{\vec{r}=\vec{r}_1,\vec{r}_2}\mu_{\vec{r}}^\alpha n_{\vec{r}}^{\alpha} \right)
\label{eq23}
\ee 
\end{widetext}
where the Lagrange multipliers $\mu$ impose the constraint that the total vector spin on the two ``orphan'' triangles equals $\vec{n}_1$ and $\vec{n}_2$. The matrix $\mathcal{M}^2$ above is a $2 \times 2$ matrix of the following form:

\begin{displaymath}
\mathcal{M}^2= 
\left( \begin{array}{cc}
\frac{\langle \vec{\phi}_{\triangle}^2(\vec{r}_1) \rangle}{3} &\frac{ \langle\vec{\phi}_{\triangle}(\vec{r}_1)\cdot \vec{\phi}_{\triangle}(\vec{r}_2) \rangle}{3}\\
\frac{ \langle\vec{\phi}_{\triangle}(\vec{r}_1)\cdot \vec{\phi}_{\triangle}(\vec{r}_2) \rangle}{3} &\frac{\langle \vec{\phi}_{\triangle}^2(\vec{r}_2) \rangle}{3}  
\end{array} \right)
\end{displaymath}
where the correlators are again calculated in the parent spin liquid state in the absence of disorder (hence, the diagonal terms are independent of $\vec{r}$). Now, it is easy to integrate out the $\vec{\mu}$ fields to obtain the effective interaction between the orphans:
\be
Z_{eff} \propto \int \mathcal{D}\vec{n}_1 \mathcal{D}\vec{n}_2 \exp \left(-\beta J_{eff} \vec{n}_1 \cdot \vec{n}_2 \right) 
\label{eq24}
\ee
where 
\be 
\beta J_{eff} \approx \frac{-\langle \vec{\phi}_{\triangle}(\vec{r}_1) \cdot \vec{\phi}_{\triangle}(\vec{r}_2)  \rangle}{\langle \vec{\phi}_{\triangle} \cdot \vec{\phi}_{\triangle} \rangle^2 }
\label{eq25}
\ee
Thus, $\beta J_{eff}$ is determined by the ``charge''-``charge'' correlator of the parent spin liquid. In
the Appendix, we use a general long-wavelength analysis of the properties of a ``Coulomb-liquid'' with
fluctuating fields to gain insight into the nature of these ``charge''-``charge'' correlations.

To summarize this section, the effects of putting non-magnetic impurities in the parent spin liquid show up in the following manner: Single vacancies leads to a spin texture that follows the spin-spin correlation (Eq ~\ref{eq15}) of the parent spin liquid. Orphan spins, generated when all but one spin are substituted by non-magnetic impurities in a simplex, however generate a texture that depends on the spin-``charge'' correlation (Eq ~\ref{eq21}) of the parent spin liquid, where the ``charge'' is located on the orphan spin simplex. The orphan spins interact via a pair-wise Heisenberg interaction $J_{eff}$ that is essentially determined by the ``charge''-``charge'' correlator (Eq ~\ref{eq25}) of the parent spin liquid, in which the two ``charges'' are located on the two orphan simplices. Note that the spin-``charge'' and ``charge''-``charge'' correlations are simply appropriate linear combinations of the spin-spin correlations. However, as we will see in the next section, their behaviour is quite different from the $1/r^2$ behaviour of the spin correlations of the pure system.

\section{Results on the pure system}
\label{sec:pure}
\subsection{Thermodynamics: Field theory and Monte-Carlo simulations}
The effective field theory results can be worked out for a pure system as the lattice is sufficiently symmetric to permit an analytical treatment despite its non-trivial seven-site basis. Working in Fourier space with this seven site basis, we obtain the following expressions for the magnetization of the kagome layer spins $m_{kag}$, and the apical spins $m_{ap}$:
\be 
m_{kag} &=& \frac{h}{2} \left(\frac{1+\frac{\rho_2 T}{2}}{3+3\rho_2 T+\rho_1 T(1+\frac{\rho_2 T}{2})} \right) \nn \\
m_{ap} &=& \frac{h}{2} \left(\frac{\rho_1 T}{3+3\rho_2 T+\rho_1 T(1+\frac{\rho_2 T}{2})} \right)
\label{eq26}
\ee
Since $\rho_1,\rho_2>0$ at any $T$, this immediately implies that $m_{kag},m_{ap} \geq 0$ for all $T$ within the large-N approximation. At $T=0$, $m_{kag}=\frac{h}{6}$ and $m_{ap} = 0$. The leading temperature corrections to this $T=0$ result are
\be
m_{kag} &=& \frac{h}{6} \left(1-\frac{T}{2}\left(\rho_2+\frac{2}{3}\rho_1 \right)+\cdots \right) \nn \\
m_{ap} &=& \frac{h}{6}\left(\rho_1 T+\cdots \right) 
\label{eq27}
\ee

Upon comparing with MC results for $O(3)$ spins (see Fig.~\ref{kagomeapical} and Fig.~\ref{sus}), we find that these effective field theory results for $m_{kag}$ and $m_{ap}$ exhibit important qualitative departure from the actual behaviour of $O(3)$ spins. First, from a comparison with
the MC data, we see that the field theory prediction
for $m_{kag}$ has the right dependence on $T$ but with the {\em wrong sign} 
of the coefficient of the leading low-temperature correction. Second
we note that our MC results demonstrate that low temperature $m_{ap}$ represents a {\em diamagnetic response} to an external field, in the sense
that the magnetization develops in a direction antiparallel to the
applied field. This diamagnetic correction is seen to be non-analytic, $m_{ap}\sim-\sqrt{T}$, and cannot be captured within our effective field theory. 

To understand these failures of the effective field theory better,
we have also studied a classical $O(4)$ Heisenberg model on the same
lattice using MC methods. From our MC results for the $O(4)$ case,
we see that $\chi_{kag}$ indeed follows the effective field theory temperature correction in Eq~\ref{eq27} at low temperature. Also, the diamagmetism of the apical layer spins is much reduced and follows $m_{ap} \sim -T$ at very low $T$. 

\begin{figure}
{\includegraphics[width=\hsize]{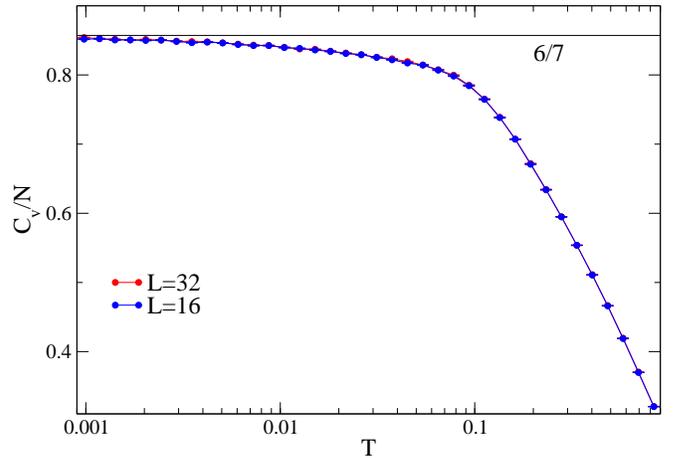}}
\caption{Specific heat $C_v$ shows no signs of a phase transition for $x=0$ pure SCGO lattice as the temperature $T$ is lowered. It converges to the value $6/7$ obtained from a mode-counting argument which shows the presence of two zero-modes per unit cell.}
\label{Cvfig}
\end{figure}

We would like to stress that a subleading correction due to thermal order-by-disorder effects is in keeping with 
the expectation from constraint-counting which is for the paramagnetic regime to persist all the way to 
$T=0$ in SCGO.\cite{chalker1} In particular, one expects SCGO O(3) spins to 
exhibit a spin liquid phase with a low-temperature specific heat of $6/7$ per spin, indicative of two zero-energy modes
per unit cell. This prediction is in quantitative agreement with our Monte Carlo results, Fig.~\ref{Cvfig}, which in addition 
shows no sign of a phase transition down to the very lowest accessed temperature.

\subsection{Dynamics: Monte-Carlo simulations}
\begin{figure}
{\includegraphics[width=\hsize]{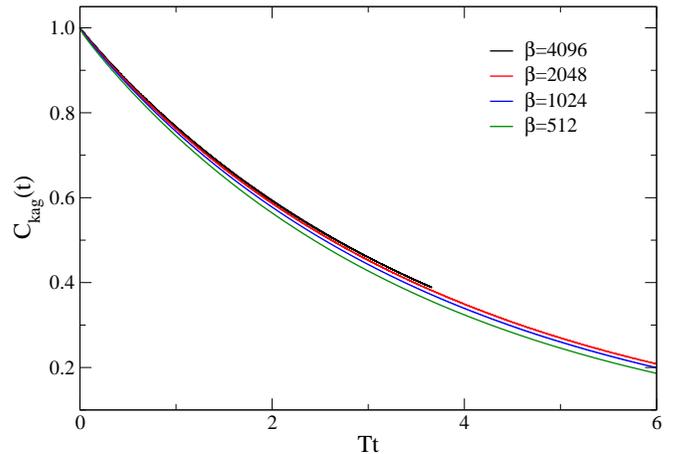}}
\caption{Autocorrelation function $C_{kag}(t) = \langle \vec{S}(t) \cdot \vec{S}(0) \rangle$ where $\vec{S}$ is a spin in the kagome layers for a system of size $L=32$ at $x=0$. Note that the autocorrelation time scales as $\frac{1}{T}$ at low $T$, which shows that the $x=0$ system remains a spin liquid down to the lowest temperatures accessed.}
\label{autocorr}
\end{figure}

\begin{figure}
{\includegraphics[width=\hsize]{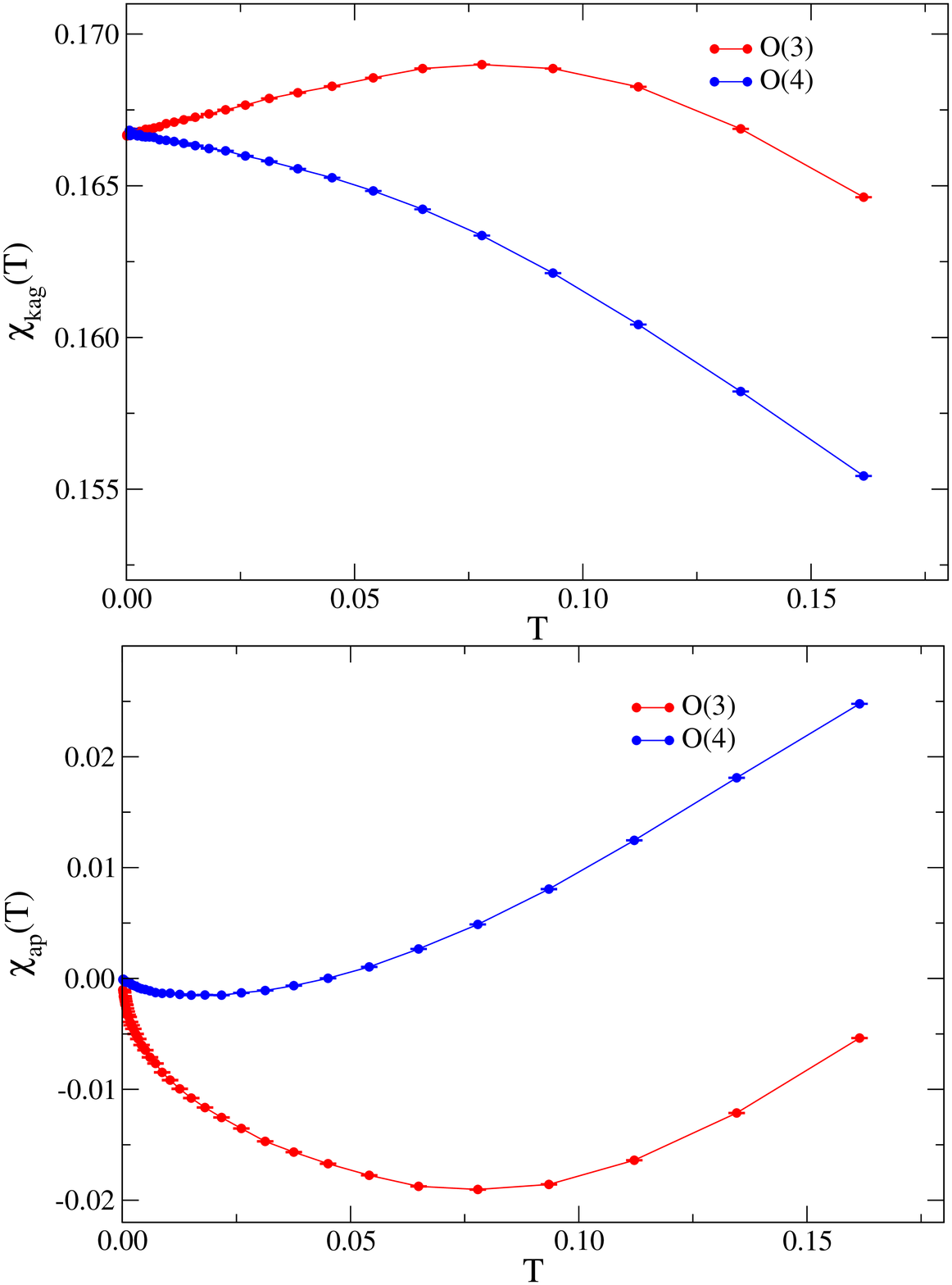}}
\caption{Zero field susceptibility of the kagome layer spins (top panel) and the apical spins (bottom panel) for $O(3)$ and $O(4)$ spins. The apical spins show a $-a\sqrt{T}$ behavior at low $T$ for the case of $O(3)$ spins. }
\label{kagomeapical}
\end{figure}

\begin{figure}
{\includegraphics[width=\hsize]{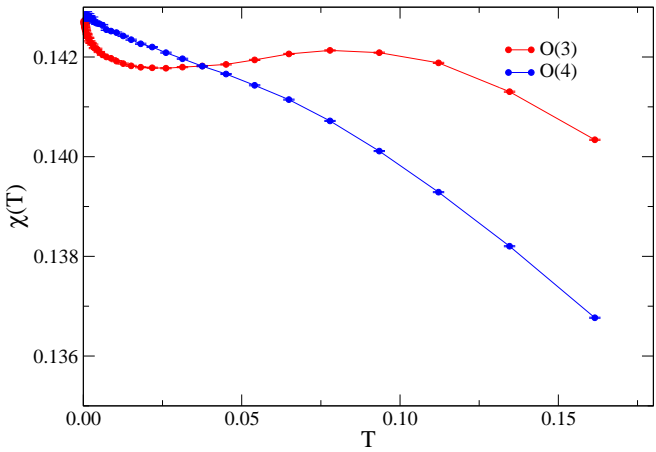}}
\caption{Total susceptibility for $O(3)$ and $O(4)$ spins. Note that the results for $O(4)$ spins
are in qualitative agreement with expectations from large-$N$ theory.}
\label{sus}
\end{figure}

A conceptually connected but nevertheless distinct diagnostic for the spin liquid regime at low
temperature consists of considering the dynamics of the system~\cite{chalker1,chalker2}, which greatly differs
between frustrated and unfrustrated systems~\cite{Keren}.
The simple prediction is for the autocorrelation function of the spins to decay exponentially, on a timescale set by the
inverse temperature. Whereas we have not done molecular dynamics simulations of the proper equations of motion including all the relevant conservation laws, our Monte Carlo results, obtained in simulations that only use strictly local single--spin-flip moves ({\em i.e.}, with the ``microcanonical''
over-relaxation moves switched off), are nonetheless strong evidence that the system
does not enter a glassy state. Instead,  the autocorrelation function $C_{kag}(t) =\frac{7}{6N} \sum_{\vec{r}} \langle \vec{S}_{\vec{r}}(t) \cdot \vec{S}_{\vec{r}}(0) \rangle$, where the sum is over spins on the two Kagome layers  exhibits an exponential decay  at a timescale which grows algebraically with the inverse temperature (Fig.~\ref{autocorr}),  in fact being precisely proportional to the inverse temperature as expected in a spin-liquid phase. Similar results were obtained
for the apical spins (not shown).

All of this taken together implies that SCGO is a model system with a spin liquid phase closely related to that
of classical Heisenberg magnets on the pyrochlore lattice, and other magnets in which thermal order-by-disorder
effects do not produce an ordered state. Furthermore, as we will see below, the effects of
vacancies are in several ways stronger in $d=2$ than in $d=3$; this, together with detailed experimental
results available in the literature, provides much of the motivation of studying SCGO as a candidate spin liquid with quenched disorder.


\section{Results on impurity effects}
\label{impurityeffects}


\subsection{A pair of vacancies on the same triangle: Single orphan physics}
To obtain the behavior of a single orphan texture, we start with the pure system and remove two magnetic sites and use the general procedure discussed
earlier to obtain the effective action of the orphan spin (Eq~\ref{eq17}). 
From this, we obtain that the orphan spin acts like a free spin a 
field $h/2$ at low $T$ and hence, $\langle n^z_{orphan} \rangle =\mathcal{B}(h/2,T)$. 
We then use this result to calculate the full texture on the lattice scale using the approach detailed in the  
 previous section and appendix. The screening of half of the magnetic field for the orphan spin at low $T$ and the detailed form of the texture obtained 
from this procedure both agree very well with the MC results for the $O(3)$ model with two vacancies on a triangular simplex; a summary of these results has already appeared in our Letter\cite{Sen_Damle_Moessner_PRL}.

Here we explore the result further by connecting it to correlations of the pure system. We cast the solution for the orphan induced spin texture in a fairly compact form using appropriate correlation functions of the parent spin liquid:
\be
&&\langle \phi^z(\vec{r}_2) \rangle =\chi(\vec{r}_2)h +\frac{\langle \phi^z(\vec{r}_1)\phi^z(\vec{r}_2) \rangle}{6(1/3-C_{nn})}\mathcal{B}(h/2,T)\nn \\  \nn\\
&+&\frac{\langle\phi^z_{\triangle}(\vec{r}_1)\phi^z(\vec{r}_2) \rangle}{\langle (\phi^z_{\triangle})^2 \rangle} \left(-\frac{3C_{nn}}{1/3-C_{nn}}\mathcal{B}(h/2,T)-3h\chi_{kag} \right) \nn \\
\label{eq28}
\ee
where $\langle \phi^z(\vec{r}_1)\phi^z(\vec{r}_2) \rangle$ is the spin-spin correlation between the spin at $\vec{r}_1$ (position of the orphan spin) and the spin at $\vec{r}_2$ in the parent spin liquid and $\langle\phi^z_{\triangle}(\vec{r}_1)\phi^z(\vec{r}_2) \rangle$ is the spin-``charge'' correlation where $\phi^z_{\triangle}(\vec{r}_1)$ is the $z$ component of the sum of the three soft-spins on the orphan simplex (in the parent spin liquid).  At low $T$ and sufficiently far away from the orphan spin, the above expression can be further simplified to give 
\be 
\langle \phi^z(\vec{r}_2) \rangle \approx \chi(\vec{r}_2)h + \frac{\langle \phi^z_{\triangle}(\vec{r}_1) \phi^z(\vec{r}_2) \rangle}{T}\mathcal{B}(h/2,T)
\label{eq29}
\ee
This latter form can also be derived directly by arguing that the internal structure of the defective simplex is unimportant for $r \gg a$ (Sec \ref{fieldtheorycomputations}).

\begin{figure}
{\includegraphics[width=\hsize]{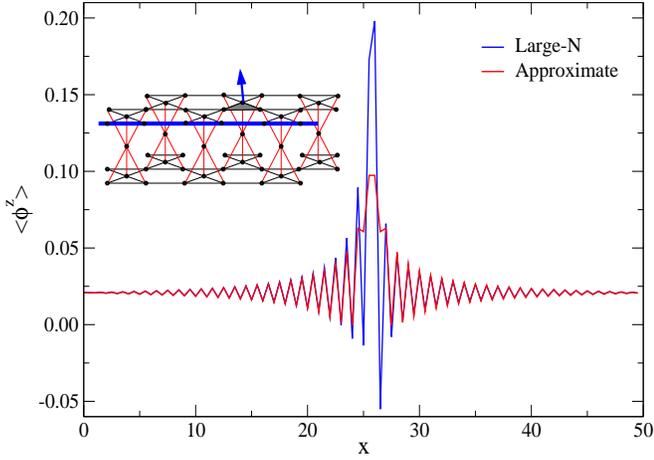}}
\caption{The texture obtained from the spin-``charge'' correlator of the parent spin liquid in the Kagome bilayer shown along a one-dimensional cut on the lattice near the orphan spin (see inset). The ``charge'' location is fixed on a triangular simplex as shown in the inset. The approximate version of the effective field theory result only deviates significantly from the full prediction (labeled ``large-$N$'' in figure) close to the orphan spin site. System size is $L=50$ at $\beta J =2048$ and $h/J=0.125$.}
\label{chargespinpattern}
\end{figure}

\begin{figure}
{\includegraphics[width=\hsize]{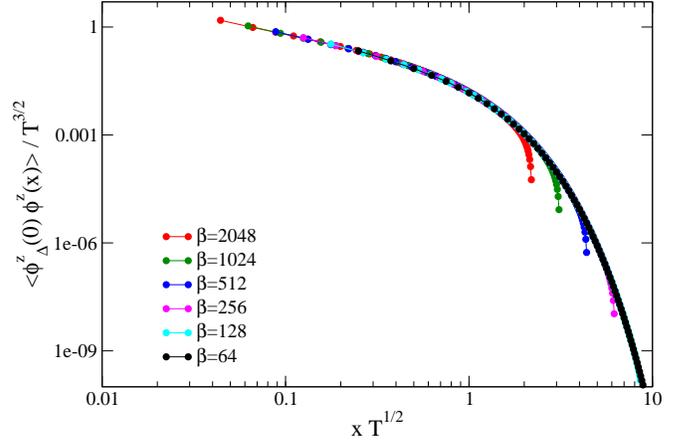}}
\caption{The spin-``charge'' correlator within our effective field theory is proportional to $T^{3/2}F_1(\vec{r}\sqrt{T})$, as expected from the Coulomb liquid analogy (see Appendix).}
\label{rinverse}
\end{figure}

The comparison between the approximate answer (Eq ~\ref{eq29}) and the
full effective field theory result (Eq ~\ref{eq28}) is shown for a system of size $L=50$ in Fig~\ref{chargespinpattern}. As is clear from the Figure, the approximation only deviates significantly from the full result when the texture very close to the orphan spin is considered.
Further, for $|\vec{r}_1 - \vec{r}_2| \gg a$ (where $a$ is the lattice spacing), the spin-``charge'' correlator $\langle \phi^{\alpha}_{\triangle}(\vec{r}_1) \phi^{\alpha}(\vec{r}_2) \rangle$
is expected to satisfy the scaling form
\begin{eqnarray} 
\langle \phi^{\alpha}_{\triangle}(\vec{r}_1) \phi^{\alpha}(\vec{r}_2) \rangle &=& \eta(\vec{r}_1)T^{3/2}F_1((\vec{r}_1-\vec{r}_2)\sqrt{T}) \; ,
\end{eqnarray}
where $F_1(\vec{x}) \sim 1/|\vec{x}|$ when $|\vec{x}| \ll 1$ and decays exponentially at large $|\vec{x}|$,
and  $\eta$ is the sublattice index of the bipartite dual lattice of simplices, taking
on a value $+1$ for the $A$-sublattice, and $-1$ for the $B$-sublattice; the rationale behind this expectation
is detailed in the appendix and relies on our analysis
 of the long-wavelength properties of a ``Coulomb liquid'' with fluctuating fields. In Fig~\ref{rinverse}, we see
that this expectation is borne out by our results.

\subsection{Form of the screening cloud}
An interesting aspect of our results is the fact
that the texture and the resulting impurity susceptibility deviates so little from the asymptotic low temperature
predictions even at sizeable temperatures of order $0.1JS^2$; this was
already noted in our earlier Letter~\cite{Sen_Damle_Moessner_PRL}. To understand
this robustness better, it is useful to consider just how
the orphan-texture complex acquires a net spin of 
$S^z_{tot}=S/2$ at $T=0$ as more and more spins around the orphan
are taken into account. 

To explore this, we begin by noting that the orphan spin is fully polarized by an infinitesimal magnetic field at $T=0$, and produces a staggered spin texture around it that decays as $1/r$ from the Coulomb phase analogy. Defining a {\it smeared} total spin operator $S^z_{tot}(\xi) = \sum_{\vec{r}} S^z(\vec{r})\exp(-r^2/\xi^2)$, where $\vec{r}$ is measured from the orphan spin site, we find that 
\be 
S^z_{tot}(\xi) &=& \frac{S}{2} + Sf(\xi) \nn \\
\mathrm{where} \mbox{~~~~} f(\xi) &\sim& 1/\xi^2, \xi \gg 1
\label{eq30}
\ee
for a two-dimensional Coulomb spin liquid (see Appendix for the calculation of this operator in the simpler case of the planar pyrochlore lattice). Thus, $S^z_{tot}(\xi)$ approaches $S/2$ quite quickly ($\xi \sim 6$ is enough to approach within $1\%$ for the planar pyrochlore lattice). 

This has important implications for the finite temperature properties. The magnetic susceptibility can be though of as arising from spin-$S/2$ orphan-texture complexes
to a given accuracy as long as the thermal correlation length $\xi(T) \sim 1/\sqrt{T}$ are larger than the length scale over which the total spin
of the zero temperature orphan-texture complex reaches $S/2$ to the same accuracy.
The robustness seen at finite temperature is thus connected with
the moment of the zero temperature orphan-texture complex approaching its asymptotic value rather
quickly in the sense of Eqn.~\ref{eq30} as one goes further and further out from the core of
this complex.

\subsection{Two orphans: Effective interactions between orphan spin textures}
\begin{figure}
{\includegraphics[width=\hsize]{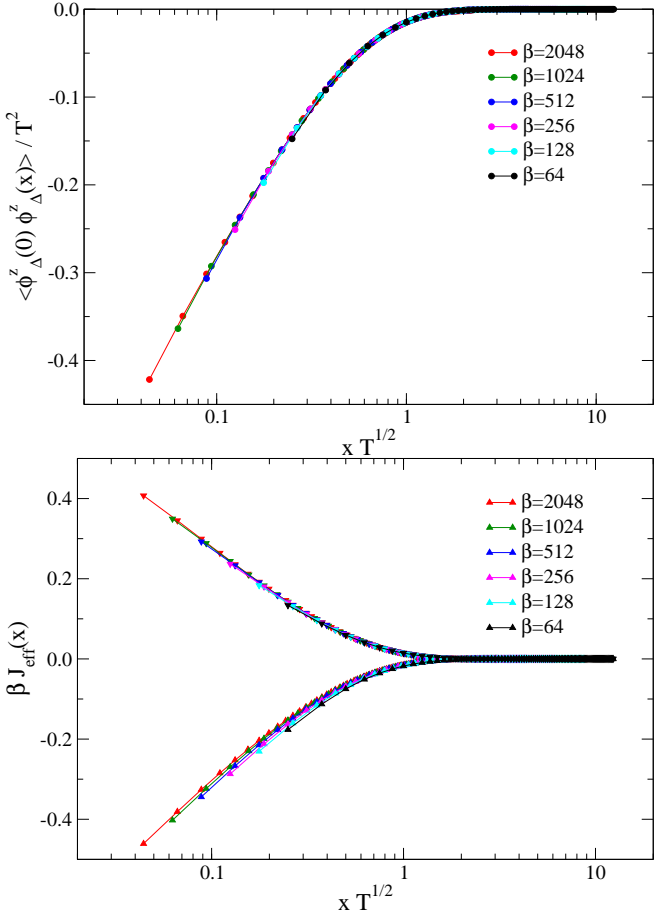}}
\caption{(Top panel)``Charge'' correlator calculated between triangular simplices on the same layer in the SCGO lattice. (Bottom panel) The effective interaction for two orphans in the same layer (upper curve) and different
layers (lower curve). System size used is $L=200$.}
\label{chargecorr}
\end{figure}

The interactions between the orphan spins can be calculated by considering the two orphans as fixed-length vectors and integrating out the rest of the soft-spin degrees of freedom $\phi(\vec{r})$ as detailed in the previous section and appendix. In this way, we obtain
\be 
J_{eff} \approx \frac{-T\langle \vec{\phi}_{\triangle}(\vec{r}_1) \cdot \vec{\phi}_{\triangle}(\vec{r}_2)  \rangle}{\langle \vec{\phi}_{\triangle} \cdot \vec{\phi}_{\triangle} \rangle^2 }
\label{eq31}
\ee
Further, for $|\vec{r}_1 - \vec{r}_2| \gg a$, where $a$ is the lattice spacing, we expect that the ``charge''-``charge'' correlator $\langle \phi^{\alpha}_{\triangle}(\vec{r}_1) \phi^{\alpha}_{\triangle}(\vec{r}_2)\rangle$ satisfies a  scaling form
\begin{eqnarray}
\langle \phi^{\alpha}_{\triangle}(\vec{r}_1) \phi^{\alpha}_{\triangle}(\vec{r}_2)\rangle &=& -\eta(\vec{r}_1) \eta(\vec{r}_2)T^2 F((\vec{r}_1-\vec{r}_2)\sqrt{T})
\end{eqnarray}
 at low temperatures, where  $\eta$ is the sublattice index of the bipartite dual lattice of simplices, taking
on a value $+1$ for the $A$-sublattice, and $-1$ for the $B$-sublattice; the rationale behind this expectation is again detailed in the appendix and relies on our analysis
 of the long-wavelength properties of a ``Coulomb liquid'' with fluctuating fields. From Fig ~\ref{chargecorr}, we
see that this expectation is borne out by our results.
Since $\langle \vec{\phi}_\triangle^2 \rangle = 3T$ at low $T$, this implies a scaling form 
\begin{eqnarray}
J_{eff}(\vec{r}_1 - \vec{r}_2,T) &=& \eta(\vec{r}_1) \eta(\vec{r}_2)T {\mathcal J}(\sqrt{T}(\vec{r}_1 - \vec{r_2}))
\end{eqnarray}

where ${\mathcal J}(y)$ is seen to have the asymptotic behaviour
\begin{eqnarray}
{\mathcal J}(\vec{y})& \sim & \log |\vec{y}| \; \; {\rm for} \; \; |\vec{y}|\ll1  \nonumber \\
{\mathcal J}(\vec{y}) &\sim& \exp(-|\vec{y}|) \; \; {\rm for} \; \; |\vec{y}| \gg 1  
\label{asymptoticformofJ}
\end{eqnarray}

The dependence on sublattice index $\eta$ leads to another interesting observation upon noting
that all triangular simplices on the upper Kagome layer have $\eta=+1$, while
all triangular simplices on the lower Kagome layer have $\eta = -1$. Therefore, two orphans in the same layer interact antiferromagnetically and have a vanishing net Curie response in the limit of low fields and temperatures smaller
than this interaction scale. On the other hand, two orphans in opposite layers couple ferromagnetically, leading to an enhanced Curie tail due to a `restituted' moment equal to that of a full free spin $S$! Note that this restituted moment arises because most of the spin-density that leads to the fractional moment of $S/2$ for a single orphan is localized close to it, as we discussed in the previous section. It is also interesting to note that this behaviour
is in sharp contrast to that of the spin textures themselves, 
which are screened at $T=0$ when the orphans are in opposite layers, since a ``charge'' zero ``dipole'' formed
by two orphans on opposite layers leads to a $1/r^2$ far-field behavior instead of the $1/r$ profile of a single spin texture at $T=0$.

\begin{figure}
{\includegraphics[width=\hsize]{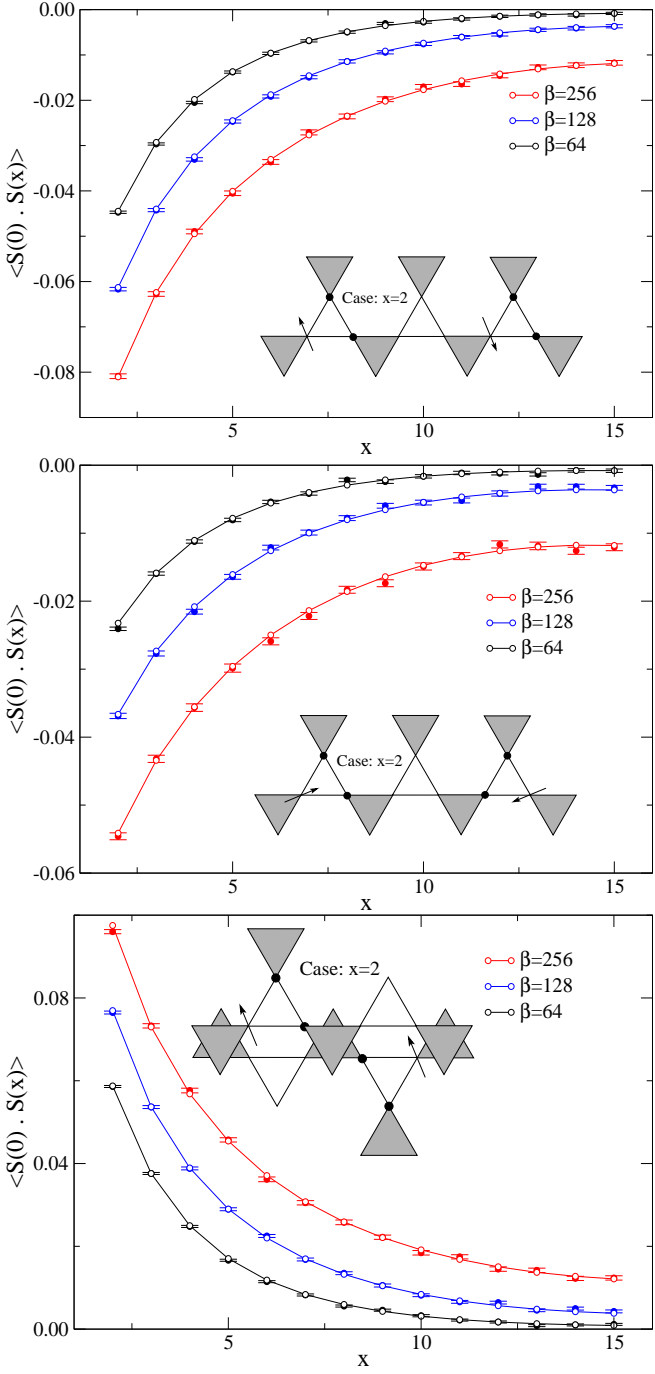}}
\caption{Agreement between effective field theory  predictions (solid lines) for orphan spin correlators, and actual results (points with error bars) for
the same quantities obtained from MC for the $O(3)$ system shown for three inequivalent orphan spin placements (shown
in corresponding insets).
}
\label{interactions1}
\end{figure}

Fig.~\ref{interactions1}  shows that the effective field theory computation for $J_{eff}(\vec{r},T)$ agrees very well with the effective interaction obtained from direct simulations of the $O(3)$ Heisenberg model with vacancies. In the simulations, we create two orphan spins by removing two vacancies each from the chosen triangular simplices. Because of the complicated geometry of the lattice, many symmetry inequivalent choices are possible and here we show three of them. We monitor $\langle \vec{S}(0) \cdot \vec{S}(\vec{r}) \rangle$ in the simulations at different $\vec{r}$ and various (low) temperatures, where $\vec{S}(0)$ and $\vec{S}(\vec{r})$ refer to the two orphan spins. This quantity is then computed using the effective field theory result for $J_{eff}(\vec{r},T)$ and the agreement is excellent in all the cases. The effective field theory computations which were done on finite lattices for SCGO, also capture the finite-size effects in the system very well.

\begin{figure}
{\includegraphics[width=\hsize]{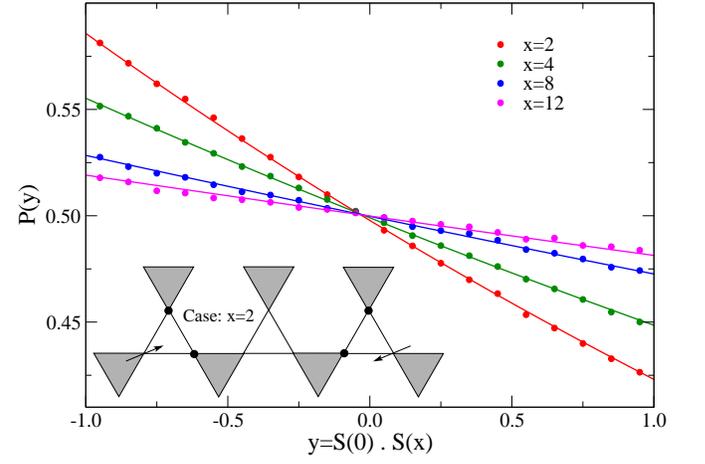}}
\caption{Probability distribution of the dot product of two orphan spins measured in MC simulations (points) exactly matches the result obtained (solid lines) by using a 
simple Heisenberg interaction term having magnitude and sign predicted by the effective field theory.}
\label{probdistwo}
\end{figure}

The probability distribution of $x = \vec{S}(0)\cdot\vec{S}(\vec{r}) $ obtained from the MC simulations (see Fig~\ref{probdistwo}) can also be fully matched to $P(x) \propto \exp(-J_{eff}(\vec{r},T)x)$ to rule out interaction terms of the form $(\vec{S}_1 \cdot \vec{S}_2)^2$ which are not forbidden on symmetry grounds, but
appear to be absent. Finally, we note that one may in principle plug this information back in
and obtain the response of the surrounding spin liquid to this pair of interacting orphan spins, and
thereby compute the low-temperature behaviour of this system of two interacting orphan-texture complexes (as emphasized
earlier in our detailed summary of the effective field theory computations).

\subsection{Three orphans: Absence of multi-spin interactions}
\begin{figure}
{\includegraphics[width=\hsize]{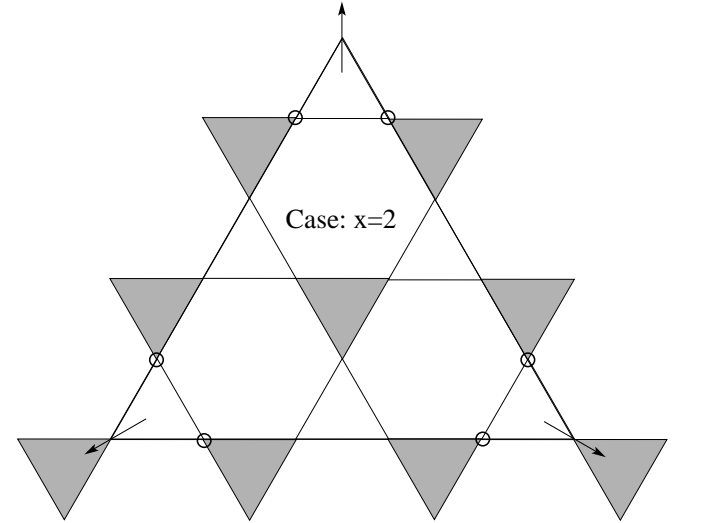}}
\caption{Three orphan spins placed in an equilateral triangle arrangement. The circles indicate the vacancies.}
\label{equilateral}
\end{figure}

From the structure of the effective action for the Lagrange-multiplier
fields detailed earlier, it is clear that our effective field theory always
gives pair-wise interactions even when the number of orphan spins is greater
than two. This is a strong prediction. The way we check this from our numerics is to place the three orphan spins in a symmetric equilateral triangle configuration as shown in Fig ~\ref{equilateral}. 
 
The first calculation we do is to calculate $\langle (\vec{S}_1+\vec{S}_2+\vec{S}_3)^2 \rangle$ for a given system size $L$ and inverse temperature $\beta$ from the effective field theory and check it with the result obtained from a pair-wise $J_{eff}(\vec{r})$ interaction (see previous section). The agreement is extremely good. We show in the table below the results from runs at $L=32$ at $\beta J=256$ for four different separations (see Fig~\ref{equilateral} for the configuration chosen in these runs). 
\begin{center}
  \begin{tabular}{| l | c | r | }
    \hline
    Seperation & MC Numerics & Large-N result \\ \hline
    02 & 2.6913(20)&  2.69270294860 \\ \hline
    04 & 2.7932(21) & 2.79425394166 \\ \hline
    08 & 2.8879(19) & 2.89057813819 \\ \hline
    12 & 2.9213(19) & 2.92543476168 \\
    \hline
  \end{tabular}
\end{center}

\begin{figure}
{\includegraphics[width=\hsize]{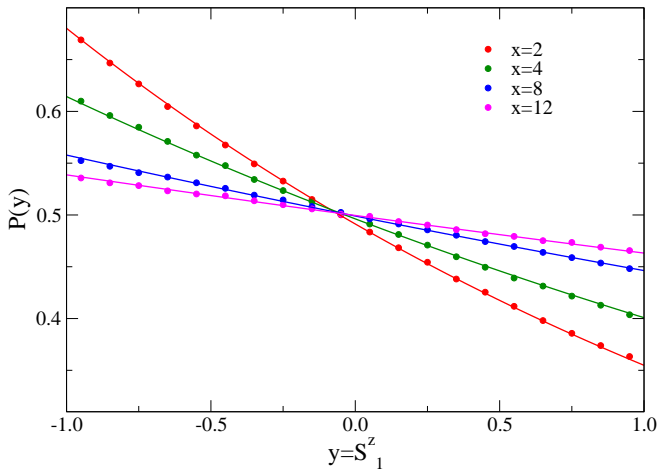}}
\caption{Probability distribution of $S_1^z$ in the three orphan problem 
when the other two orphans are fixed in the $\hat{z}$ direction (as detailed in text). Points
represent values measured in MC simulations, while lines are fits obtained from the effective field
theory prediction of purely bilinear two-body exchange interactions. }
\label{probdisthree}
\end{figure}

Secondly, we probe the probability distribution of the orphan spins to see how well a pair-wise interaction picture can explain it. Since, the full probability distribution function is complicated even for pair-wise interactions, we fix two of the orphan spins $\vec{S}_2=\vec{S}_3=\hat{z}$ in the MC numerics and then monitor the probability distribution $S_1^z$ of the third orphan spin. If the pair-wise interaction picture is true, then the resulting distribution will be proportional to $\exp(-2J_{eff}(\vec{r})S_1^z)$, where $J_{eff}(\vec{r})$ is the pair-wise interaction strength, and this is exactly what we observe from the simulations (Fig ~\ref{probdisthree}). 

\section{Generalisations}
\subsection{Orphan tetrahedra}
The above arguments all hold for orphan (thrice-defective) tetrahedra as well as the orphan (twice-defective) triangles discussed throughout. The emergent gauge charge of an orphan tetrahedron is opposite to that of a triangle in the same layer. In a random dilution model, the probability of the former is $3x^2(1-x)$ while that of the latter is $4x^3(1-x)$, much smaller in the limit of small $x$.

\subsection{Other lattices, and dimensions}
Several central results readily generalise to other lattices: orphan spins can be induced in O$(n \geq 4)$ spin liquids on the kagome lattice, and for O$(n \geq 3)$ spin liquids on the pyrochlore lattice, which is of course where they were first identified \cite{Moessner_berlinsky,Henley_2000}. Carrying emergent gauge charges, they interact via a $r^{-d+2}$ Coulomb interaction for $d \geq 3$, the generalisation of the $d=2$ logarithm implied by Eq. \ref{asymptoticformofJ}; analogously, their textures decay as $r^{-d+1}$.

\section{Conclusions and outlook}
Motivated by the extensive set of experimental data available for SCGO, we have studied the Heisenberg model on the corresponding lattice as a model system for a spin liquid. We have established in detail that  SCGO remains in a spin liquid phase down to the lowest temperatures, as expectation based from mode counting
arguments.

We have then studied in detail the response of this spin liquid to the inclusion of disorder in the form of vacancies. By developing a field theory which captures the hard-spin nature of spins near a vacancy and treats entropic effects
on the same footing as energetics, we were able to get an analytical handle on the resulting phenomena---these predictions were found to be in excellent
agreement with direct Monte-Carlo simulations of the Heisenberg model with vacancies.

In particular, we have found that vacancies that leave behind more than two spins in a simplex generically lead to a regular low-temperature limit of the susceptibility. On the other hand, the presence of an orphan spin, all of whose neighbouring spins in a simplex have been removed by dilution, leads to a Curie tail in the susceptibility, with a characteristic $1/T$ divergence in
the low temperature limit,  corresponding to the susceptibility of a free
``spin $S/2$''. This fractional moment occurs as a combination of two effects: First,
the coupling to
the surrounding spin liquid ``screens out'' half of the external field $h$ seen
by the orphan spin, so that it behaves as a spin $S$ in a field $h/2$. And second, the surrounding spin texture develops
a diamagnetic response that ``cancels off'' half the polarization of the orphan
spin. 

These orphan-texture complexes experience long-range mutual interactions on account of their extended nature, captured intuitively by an analogy to the electrostatics of the Coulomb phase of the field theory describing the spin liquid. One of the central advances in this work is our derivation of the relevant scaling functions fully describing vacancies in SCGO---these quantitatively capture both thermal and energetic effects on an equal footing, requiring only knowledge of the correlators of the pure system!

The field of vacancies in unconventional magnetic states is a rich one with a long and interesting history, dating back (at least) all the way back to Villain's work on canted spin states in his seminal paper on insulating spin glasses~\cite{Villian}.
This remains an exciting frontier, with many interesting avenues worth exploring. The obvious next step would focus on the many-body states resulting from the orphan interactions described here.  Understanding this many-body
physics is of considerable general interest, since
the form of our interactions is actually a quite general aspect
of defects which cause violations of the emergent Gauss law constraint
in Coulomb phases. On general grounds, with such defects randomly distributed, this interaction can naturally lead to the appearance of a spin-glass phase. We are currently investigating this issue in detail~\cite{Damle_etal}. Returning to the specific case of SCGO, it will then remain to be seen whether the spin glass transition observed in experiments can be related to orphan-texture freezing.

More broadly for the case of SCGO, we hope that our work will stimulate a more detailed study with the aim of better characterising the disorder present there, given that our analysis of the NMR lineshapes\cite{Sen_Damle_Moessner_PRL} has found that the Curie tail cannot be explained with reference to orphans induced by the nominal amount of uncorrelated vacancies in this series of compounds. In this context,
it would be useful to characterize in more detail the
statistics of substitution of Ga on the Cr sites, including possible
correlations, as well as better characterize other forms of disorder
that may be playing an equally important role in the temperature regime
studied experimentally.

\section{Acknowledgements}
We gratefully acknowledge useful discussions with  J.~Chalker, A.~Das,  D.~Dhar, 
C.~Henley, P.~Mendels, and A.~W.~Sandvik, 
funding from DST SR/S2/RJN-25/2006 (KD) and IFCPAR/CEFIPRA Project 4504-1 (KD),
financial support for collaborative visits from the Fell Fund (Oxford),
the ICTS TIFR (Mumbai), ARCUS (Orsay) and MPIPKS (Dresden), as well as
computational resources at MPIPKS and TIFR.

\appendix
\begin{widetext}
\section{Long-wavelength description of ``charge''-``charge'' and spin-``charge'' correlators}
Our results rely crucially on the ``charge''-``charge'' and spin-``charge''
correlators satisfying the scaling forms
\begin{eqnarray}
\langle \phi^{\alpha}_{\triangle}(\vec{r}_1) \phi^{\alpha}_{i}(\vec{r}_2) \rangle &=& \eta(\vec{r}_1)T^{3/2}F_1((\vec{r}_1-\vec{r}_2)\sqrt{T}) \,, \nonumber \\
\langle \phi^{\alpha}_{\triangle}(\vec{r}_1) \phi^{\alpha}_{\triangle}(\vec{r}_2)\rangle &=&-\eta(\vec{r_1})\eta(\vec{r}_2)T^2 F((\vec{r}_1-\vec{r}_2)\sqrt{T})
\end{eqnarray}
where $\alpha$ denotes the spin space index of $\phi$, and by attaching
a real-space index $i$, we
have emphasized that $\phi$ is best thought of as a spatial
vector with orientation given by that of the corresponding dual lattice link on which this field lives. In the
above, $\eta$ is the sublattice index of the bipartite dual lattice of simplices, taking
on a value $+1$ for the $A$-sublattice, and $-1$ for the $B$-sublattice. 

These scaling forms are a consequence of the fact
that $\phi^{\alpha}$ behaves at low temperature like a fluctuating
magnetic field on the links of the dual lattice, while $\phi^{\alpha}_{\triangle}$ is proportional to the divergence of this fluctuating magnetic field, {\em i.e.} the fluctuating
magnetic ``charge'' on a simplex .
To understand this scaling behaviour, it is useful
to consider a coarse-grained theory formulated directly in the continuum.
The success of this continuum approach relies on the fact that
the geometric details of the lattice  only affect the short-distance
form of these correlators (at scales $r \sim a$, where $a$ is the lattice spacing).
The continuum theory detailed below is therefore expected to apply in a
broad regime $|\vec{r}|/a \gg 1$ with no restrictions on $|\vec{r}|/\xi(T) \sim |\vec{r}|\sqrt{T}$.

With this in mind, we start with the continuum effective action
for a fluctuating classical magnetic field
\begin{eqnarray}
{\mathcal S} & = & \frac{\kappa}{2}\int_{\Lambda} d^dx (\vec{B^{\alpha}})^2 + \frac{\Delta_c}{2T}\int_{\Lambda} d^d x (\nabla \cdot \vec{B^{\alpha}})^2
\end{eqnarray}
where the repeated spin-space vector index $\alpha$ is summed over,
and the subscript $\Lambda$ serves to remind us that
this theory is defined with an upper-cutoff $\Lambda \sim 1/a$ in momentum
space.
In the above, the first term is entropic in origin (and hence, it
tends to a finite value as $T \rightarrow 0$), 
while
the second represents the Boltzmann weight for creating magnetic
``charges'', ${\Delta_c}$ and $\kappa$ are phenomenological constants related
to the energy barrier for producing ``charges'' and the ``magnetic permeability'' of the medium,
and we identify
\begin{eqnarray}
\phi^{\alpha}_{i} \sim B^{\alpha}_{i}
\end{eqnarray}
with sign chosen so that the magnetic field points from a $A$-sublattice simplex to a $B$-sublattice simplex
in the microscopic version of the theory.

In order to work with this continuum action, we
decompose the magnetic field into a curl-free pure gradient
part, and a gradient free purely rotational part
\begin{eqnarray}
\vec{B^{\alpha}} &\sim & (\hat{z} \times \nabla) {\mathfrak a}^{\alpha} - \nabla {\mathfrak u}^{\alpha} 
\end{eqnarray}
and perform the functional integrals over ${\mathfrak a}$ and ${\mathfrak u}$.
Since all correlations considered here
are diagonal in the spin space index $\alpha$, which plays no role below,
we drop it in the rest of this discussion.

We now have the correspondence
\begin{eqnarray}
\phi_{\XBox} (\vec{r}) \; {\rm or} \; \phi_{\triangle}(\vec{r}) & \sim \; \eta(\vec{r}) \nabla \cdot \vec{B} \; \sim  & -\eta(\vec{r}) \nabla^2 {\mathfrak u}
\end{eqnarray}
where $\eta$ is $+1$ ($-1$) if $\phi_{\XBox}(\vec{r})$ or $\phi_{\triangle}(\vec{r})$ represents
the total spin of a $A$-sublattice ($B$-sublattice) simplex at $\vec{r}$; thus, lattice scale distinctions only enter our theory through $\eta(\vec{r})$.

For the ``charge''-``charge'' correlators, this immediately gives in $d=2$
\begin{eqnarray}
\langle \phi_{\XBox} (\vec{r}_1) \phi_{\XBox} (\vec{r}_2) \rangle \; {\rm or} \; \langle \phi^{\alpha}_{\triangle}(\vec{r}_1) \phi^{\alpha}_{\triangle}(\vec{r}_2)\rangle  & \sim & T \eta(\vec{r}_1) \eta(\vec{r}_2)\int^{\Lambda} d^2 q \frac{\vec{q}^2\exp(i \vec{q} \cdot (\vec{r}_1 - \vec{r}_2))}{\Delta_c\vec{q}^2 + \kappa T}
\end{eqnarray}
which, apart from an extremely short-ranged part [represented in our continuum approach as a contribution proportional to $\delta_{\Lambda}^d(\vec{r} - \vec{r}^{'}) \equiv \int^{\Lambda} Td^d q \exp(i \vec{q} \cdot (\vec{r} - \vec{r}^{'}))$], reduces to
\begin{eqnarray}
\langle \phi_{\XBox} (\vec{r}_1) \phi_{\XBox} (\vec{r}_2) \rangle \; {\rm or} \; \langle \phi^{\alpha}_{\triangle}(\vec{r}_1) \phi^{\alpha}_{\triangle}(\vec{r}_2)\rangle  &\sim & - \eta(\vec{r}_1) \eta(\vec{r}_2) T^2 \int^{\Lambda/\sqrt{T}} d^2 q \frac{\exp(i \vec{q} \cdot (\vec{r}_1 - \vec{r}_2)\sqrt{T})}{\Delta_c\vec{q}^2 + \kappa}
\end{eqnarray}
which clearly yields the scaling form mentioned above.
Furthermore, from the structure of the integral in the above, it is clear
that $F(x)$ has precisely the small and large $x$ behaviour
described in the main text.

For the ``charge''-spin correlator, we have in $d=2$
\begin{eqnarray}
\langle \phi^{\alpha}_{\triangle}(\vec{r}_1) \phi^{\alpha}_{i}(\vec{r}_2) \rangle & \sim & \eta(\vec{r}_1) T^{3/2}\int^{\Lambda/\sqrt{T}} d^2 q  \frac{q_{i} \exp(i \vec{q} \cdot (\vec{r}_1 - \vec{r}_2)\sqrt{T})}{\Delta_c\vec{q}^2 + \kappa}
\end{eqnarray}
which immediately implies the scaling behaviour quoted at the outset
of this appendix.

\section{Details of the orphan spin texture calculations}
\label{fieldtheoryappendix}
The matrix describing the action (see Eq ~\ref{eq4}) for the pure system (including Lagrange multipliers and a  shift in the zero of energy to set the ground state energy to zero) is Fourier transformed to read
\begin{displaymath}
\mathcal{M}(\mathbf{k})= 
\left( \begin{array}{ccccccc}
1+\frac{\rho_1}{2\beta J} & \frac{1}{2}(1+e^{ik_y}) & \frac{1}{2}(1+e^{ik_x}) & 0 & 0 & 0 & \frac{1}{2} \\
 \frac{1}{2}(1+e^{-ik_y}) & 1+\frac{\rho_1}{2\beta J} & \frac{1}{2}(1+e^{i(k_x-k_y)}) & 0 & 0 & 0 & \frac{1}{2} \\
 \frac{1}{2}(1+e^{-ik_x}) & \frac{1}{2}(1+e^{-i(k_x-k_y)}) & 1+\frac{\rho_1}{2\beta J} & 0 & 0 & 0 & \frac{1}{2} \\
 0 & 0 & 0 & 1+\frac{\rho_1}{2\beta J} &\frac{1}{2}(1+e^{-ik_y}) & \frac{1}{2}(1+e^{-ik_x}) & \frac{1}{2} \\
 0 & 0 & 0 & \frac{1}{2}(1+e^{ik_y}) & 1+\frac{\rho_1}{2\beta J} & \frac{1}{2}(1+e^{-i(k_x-k_y)}) & \frac{1}{2} \\
 0 & 0 & 0 & \frac{1}{2}(1+e^{ik_x}) & \frac{1}{2}(1+e^{i(k_x-k_y)}) & 1+\frac{\rho_1}{2\beta J} & \frac{1}{2} \\
 \frac{1}{2} & \frac{1}{2} & \frac{1}{2} & \frac{1}{2} & \frac{1}{2} & \frac{1}{2}& 1+\frac{\rho_2}{2\beta J}
\end{array} \right)
\end{displaymath}
See Fig ~\ref{Fig1_lattice} for the numbering convention chosen for the $7$-point unit cell and $\hat{x}$ and $\hat{y}$ vectors for the Bravais lattice. The Fourier transform is defined as $\phi_{\alpha}(\vec{r}) =\frac{1}{\sqrt{N_d}}\sum_{\vec{k}}\exp(-i\vec{k}\cdot \vec{r})\phi_{\alpha}(\vec{k})$ where $\alpha=0,1,\cdots,6$ is the sublattice index and these are given the coordinate $\vec{r}$ of the direct Bravais lattice point, and $N_d=L^2$ is the number of sites in the Bravais lattice. We denote the unitary matrix that diagonalizes $M(\mathbf{k})$ by $U(\mathbf{k})$: $U(\mathbf{k})^{\dagger} M(\mathbf{k}) U(\mathbf{k}) = D(\mathbf{k})$, where $D(\mathbf{k})$ is a diagonal matrix. Then, going to the variables $\sigma_{\alpha}(\vec{k})$, where $\phi_{\alpha}(\vec{k}) = \sum_{\beta}U_{\alpha \beta}(\vec{k})\sigma_{\beta}(\vec{k})$ diagonalizes the quadratic interaction matrix.

For calculating the orphan spin texture, we impose three constraints in the soft-spin calculation as explained in Section~\ref{vacancies}, two for the two spins being removed, and one that enforces the condition that $\langle n^z_{orphan} \rangle =\mathcal{B}(h/2,T)$ as a hard constraint. Furthurmore, since we are interested in calculating the full texture, we introduce a ``source field'' $J_{\alpha}(\vec{r})$ which couples linearly to $\phi_{\alpha}^z(\vec{r})$.  Going to momentum space, and changing to $\sigma_{\alpha}(\vec{k})$ variables then leads to the following path integral (for the $z$ component, parallel to the direction of the external uniform magnetic field $h$)
\be 
\mathcal{Z} &=& \int D\sigma_{\alpha}^R(\vec{k})D\sigma_{\alpha}^I(\vec{k}) D\mu_3 D\lambda_4 D\lambda_5 \exp[-\beta J \sum_{\vec{k},\alpha}D_{\alpha \alpha}(\vec{k})((\sigma_{\alpha}^R(\vec{k}))^2+(\sigma_{\alpha}^I(\vec{k}))^2)+\beta h \sqrt{N_d}\sum_{\alpha,\beta}U^R_{\beta \alpha}(\vec{k}=0)\sigma^R_{\alpha}(\vec{k}=0) \nn \\
&+&\sum_{\vec{k},\alpha}(\tilde{J}^R_{\alpha}(\vec{k})\sigma^R_{\alpha}(\vec{k})+\tilde{J}^I_{\alpha}(\vec{k})\sigma^I_{\alpha}(\vec{k}))+i\sum_{\vec{k},\alpha}(\mu_3 \tilde{U}_{e,3}(\vec{r}_0,\vec{k},\alpha)+\lambda_4 \tilde{U}_{e,4}(\vec{r}_0,\vec{k},\alpha)+\lambda_5 \tilde{U}_{e,5}(\vec{r}_0,\vec{k},\alpha))\sigma^R_{\alpha}(\vec{k}) \nn \\
&+&i\sum_{\vec{k},\alpha}(\mu_3 \tilde{U}_{o,3}(\vec{r}_0,\vec{k},\alpha)+\lambda_4 \tilde{U}_{o,4}(\vec{r}_0,\vec{k},\alpha)+\lambda_5 \tilde{U}_{o,5}(\vec{r}_0,\vec{k},\alpha))\sigma^I_{\alpha}(\vec{k}) - i \sqrt{N_d}\mu_3 \mathcal{B}(h/2,T)]
\ee

where we assumed that the orphan spin lives on sublattice $3$. $\mu_3$ imposes the length constraint on the orphan spin and $\lambda_4$ and $\lambda_5$ impose $\phi^z_{\alpha}(\vec{r})=0$ for the two spins removed from the orphan simplex. Furthermore, $J_{\alpha}(\vec{k}) = \sum_{\beta}U_{\alpha \beta}(\vec{k})\tilde{J}_{\beta}(\vec{k})$ and 
\be 
\tilde{U}_{e,3}(\vec{r}_0,\vec{k},\alpha)&=&U^{R}_{3\alpha}(\vec{k})\cos(\vec{k} \cdot \vec{r}_0)+U^{I}_{3\alpha}(\vec{k})\sin(\vec{k} \cdot \vec{r}_0) \nn \\
\tilde{U}_{e,4}(\vec{r}_0,\vec{k},\alpha)&=&U^{R}_{4\alpha}(\vec{k})\cos(\vec{k} \cdot \vec{r}_0+k_y)+U^{I}_{4\alpha}(\vec{k})\sin(\vec{k} \cdot \vec{r}_0+k_y) \nn \\
\tilde{U}_{e,5}(\vec{r}_0,\vec{k},\alpha)&=&U^{R}_{5\alpha}(\vec{k})\cos(\vec{k} \cdot \vec{r}_0+k_x)+U^{I}_{5\alpha}(\vec{k})\sin(\vec{k} \cdot \vec{r}_0+k_x) \nn \\
\tilde{U}_{o,3}(\vec{r}_0,\vec{k},\alpha)&=&U^{R}_{3\alpha}(\vec{k})\sin(\vec{k} \cdot \vec{r}_0)-U^{I}_{3\alpha}(\vec{k})\cos(\vec{k} \cdot \vec{r}_0) \nn \\
\tilde{U}_{o,4}(\vec{r}_0,\vec{k},\alpha)&=&U^{R}_{4\alpha}(\vec{k})\sin(\vec{k} \cdot \vec{r}_0+k_y)-U^{I}_{4\alpha}(\vec{k})\cos(\vec{k} \cdot \vec{r}_0+k_y) \nn \\
\tilde{U}_{o,5}(\vec{r}_0,\vec{k},\alpha)&=&U^{R}_{5\alpha}(\vec{k})\sin(\vec{k} \cdot \vec{r}_0+k_x)-U^{I}_{5\alpha}(\vec{k})\cos(\vec{k} \cdot \vec{r}_0+k_x)
\ee
Solving the above path integral to obtain $\mathcal{Z}({\tilde{J}})$, the spin texture can then be calculated by evaluating 
\be
\langle \sigma^R_{\alpha}(\vec{k})\rangle &=& lim_{\tilde{J} \rightarrow 0} \frac{1}{2\mathcal{Z}({\tilde{J}})} \frac{\partial\mathcal{Z}({\tilde{J}})}{\partial \tilde{J}^R_{\alpha}(\vec{k})} \nn \\
\langle \sigma^I_{\alpha}(\vec{k})\rangle &=& lim_{\tilde{J} \rightarrow 0} \frac{1}{2\mathcal{Z}({\tilde{J}})} \frac{\partial\mathcal{Z}({\tilde{J}})}{\partial \tilde{J}^I_{\alpha}(\vec{k})}
\ee
The field $\langle \phi^z_{\alpha}(\vec{r})\rangle$ can then be obtained from $\langle \sigma^R_{\alpha}(\vec{k})\rangle$ and $\langle \sigma^I_{\alpha}(\vec{k})\rangle$ which leads to the result displayed in Eq~\ref{eq28}.

\section{Details of field theoretic calculation of interactions between orphans}
\label{app:orphint}

Here, we show some essential steps needed to obtain the effective interaction between two orphan spins. As discussed in Section~\ref{orphans}, we keep the orphan spins as unit vectors and intergate the rest of the soft-spins. Let us consider the specific case when the orphan spins are both placed on the sublattice $\alpha=3$ (other cases can be similarly considered). Then we impose six constraints, four for the four spins being removed and two for pinning the two orphan spins to be $n^z_1$ and $n^z_2$. We will need to remember that the $(n^z_1)^2$ and $(n^z_2)^2$ terms in the action can be combined with similar  $x$ and $y$ terms and are unimportant since these are unit vectors. The only terms that will be generated in the effective action will be of the form $h s_1 n^z_{1}$, $h s_2 n^z_{2}$ and $J_{eff} n^z_{1} n^z_{2}$ (the corresponding $x$ and $y$ terms will combine with this to give the $J_{eff} \vec{n}_1 \cdot \vec{n}_2$). The full path integral (for the $z$ part) is of the following form:
\be 
\mathcal{Z} &=& \int D\sigma_{\alpha}^R(\vec{k})D\sigma_{\alpha}^I(\vec{k}) D\mu_1 D\lambda_2 D\lambda_3 D\mu_4 D\lambda_5 D\lambda_6 \exp[-\beta J \sum_{\vec{k},\alpha}D_{\alpha \alpha}(\vec{k})((\sigma_{\alpha}^R(\vec{k}))^2+(\sigma_{\alpha}^I(\vec{k}))^2)-i\mu_1 \sqrt{N_d}n_{1}^{z}-i\mu_4\sqrt{N_d}n_{2}^{z}  \nn \\
&+&\beta h \sqrt{N_d}\sum_{\alpha,\beta}U^R_{\beta \alpha}(\vec{k}=0)\sigma^R_{\alpha}(\vec{k}=0)+ i\mu_1\sum_{\vec{k},\alpha}(\tilde{U}_{e,3}(\vec{r}_1,\vec{k},\alpha)\sigma_{\alpha}^R(\vec{k})+\tilde{U}_{o,3}(\vec{r}_1,\vec{k},\alpha)\sigma_{\alpha}^I(\vec{k})) \nn \\
&+& i\lambda_2\sum_{\vec{k},\alpha}(\tilde{U}_{e,4}(\vec{r}_1,\vec{k},\alpha)\sigma_{\alpha}^R(\vec{k})+\tilde{U}_{o,4}(\vec{r}_1,\vec{k},\alpha)\sigma_{\alpha}^I(\vec{k}))+  i\lambda_3\sum_{\vec{k},\alpha}(\tilde{U}_{e,5}(\vec{r}_1,\vec{k},\alpha)\sigma_{\alpha}^R(\vec{k})+\tilde{U}_{o,5}(\vec{r}_1,\vec{k},\alpha)\sigma_{\alpha}^I(\vec{k}))\nn \\
&+& i\mu_4\sum_{\vec{k},\alpha}(\tilde{U}_{e,3}(\vec{r}_2,\vec{k},\alpha)\sigma_{\alpha}^R(\vec{k})+\tilde{U}_{o,3}(\vec{r}_2,\vec{k},\alpha)\sigma_{\alpha}^I(\vec{k}))+ i\lambda_5\sum_{\vec{k},\alpha}(\tilde{U}_{e,4}(\vec{r}_2,\vec{k},\alpha)\sigma_{\alpha}^R(\vec{k})+\tilde{U}_{o,4}(\vec{r}_2,\vec{k},\alpha)\sigma_{\alpha}^I(\vec{k}))\nn \\
&+& i\lambda_6\sum_{\vec{k},\alpha}(\tilde{U}_{e,5}(\vec{r}_2,\vec{k},\alpha)\sigma_{\alpha}^R(\vec{k})+\tilde{U}_{o,5}(\vec{r}_2,\vec{k},\alpha)\sigma_{\alpha}^I(\vec{k}))]
\ee  

Integrating out the $\sigma^{R,I}$ fields from the above path integral gives the following:
\be 
\mathcal{Z} \propto \int D \mu_1 D\lambda_2 D\lambda_3 D\mu_4 D\lambda_5 D\lambda_6 \exp \left(-N_d\lambda^T M \lambda +i \sqrt{N_d}\sum_{\alpha=2,3,5,6}c_\alpha \lambda_\alpha+i\sqrt{N_d}\sum_{\alpha=1,4}(c_\alpha -n^z_\alpha)\mu_\alpha \right)  
\ee
where the interaction matrix $2M$ is defined below:
\begin{displaymath}
\left( \begin{array}{cccccc}
\langle \phi^z_{3}(\vec{r}_1)^2\rangle & \langle  \phi_{3}^z(\vec{r}_1) \phi_{4}^z(\vec{r}_1) \rangle & \langle  \phi_{3}^z(\vec{r}_1) \phi_{5}^z(\vec{r}_1) \rangle & \langle \phi_{3}^z(\vec{r}_1) \phi_{3}^z(\vec{r}_2) \rangle & \langle \phi_{3}^z(\vec{r}_1) \phi_{4}^z(\vec{r}_2) \rangle & \langle \phi_{3}^z(\vec{r}_1) \phi_{5}^z(\vec{r}_2) \rangle  \\
\langle \phi_{3}^z(\vec{r}_1)\phi_{4}^z(\vec{r}_1)\rangle & \langle \phi_{4}^z(\vec{r}_1)^2 \rangle & \langle \phi_{4}^z(\vec{r}_1)\phi_{5}^z(\vec{r}_1) \rangle & \langle \phi_{4}^z(\vec{r}_1) \phi_{3}^z(\vec{r}_2 \rangle & \langle \phi_{4}^z(\vec{r}_1) \phi_{4}^z(\vec{r}_2) \rangle & \langle \phi_{4}^z(\vec{r}_1) \phi_{5}^z(\vec{r}_2) \rangle  \\
\langle \phi_{3}^z(\vec{r}_1)\phi_{5}^z(\vec{r}_1)\rangle & \langle \phi_{4}^z(\vec{r}_1) \phi_{5}^z(\vec{r}_1) \rangle & \langle \phi_{5}^z(\vec{r}_1)^2\rangle & \langle \phi_{5}^z(\vec{r}_1) \phi_{3}^z(\vec{r}_2) \rangle & \langle \phi_{5}^z(\vec{r}_1) \phi_{4}^z(\vec{r}_2) \rangle & \langle \phi_{5}^z(\vec{r}_1) \phi_{5}^z(\vec{r}_2) \rangle  \\
\langle \phi_{3}^z(\vec{r}_1)\phi_{3}^z(\vec{r}_2)\rangle & \langle \phi_{4}^z(\vec{r}_1) \phi_{3}^z(\vec{r}_2) \rangle & \langle \phi_{5}^z(\vec{r}_1) \phi_{3}^z(\vec{r}_2)\rangle & \langle \phi_{3}^z(\vec{r}_2)^2 \rangle & \langle \phi_{3}^z(\vec{r}_2) \phi_{4}^z(\vec{r}_2) \rangle & \langle \phi_{3}^z(\vec{r}_2) \phi_{5}^z(\vec{r}_2) \rangle  \\
\langle \phi_{3}^z(\vec{r}_1)\phi_{4}^z(\vec{r}_2)\rangle & \langle \phi_{4}^z(\vec{r}_1) \phi_{4}^z(\vec{r}_2) \rangle & \langle \phi_{5}^z(\vec{r}_1) \phi_{4}^z(\vec{r}_2)\rangle & \langle \phi_{3}^z(\vec{r}_2) \phi_{4}^z(\vec{r}_2) \rangle & \langle \phi_{4}^z(\vec{r}_2)^2 \rangle & \langle \phi_{4}^z(\vec{r}_2) \phi_{5}^z(\vec{r}_2) \rangle  \\
\langle \phi_{3}^z(\vec{r}_1)\phi_{5}^z(\vec{r}_2)\rangle & \langle \phi_{4}^z(\vec{r}_1) \phi_{5}^z(\vec{r}_2) \rangle & \langle \phi_{5}^z(\vec{r}_1) \phi_{5}^z(\vec{r}_2)\rangle & \langle \phi_{3}^z(\vec{r}_2) \phi_{5}^z(\vec{r}_2) \rangle & \langle \phi_{4}^z(\vec{r}_2) \phi_{5}^z(\vec{r}_2) \rangle & \langle \phi_{5}^z(\vec{r}_2)^2 \rangle 
\end{array} \right)
\end{displaymath}
 and the coefficients $c_\alpha = h\chi_{kag}$ ($h\chi(\vec{r})$ in general). The above matrix is the $6 \times 6$ matrix $P_{\Lambda}CP_{\Lambda}$ for the two orphan case. The correlators in the matrix $M$ are calculated in the parent spin liquid state (i.e. no disorder) in a zero magnetic field.

A numerically simple method to solve the remaining $\lambda$ and $\mu$ integrals is by going to the diagonal basis for $M$. Thus, we use (below we use the notation $\mu_1=\lambda_1$ and $\mu_4=\lambda_4$ for brevity)
\be 
V^TMV = \bar{M}, \mbox{~~~}VV^T=\mathcal{I}, \mbox{~~~~~}\lambda_\alpha=V_{\alpha \beta}\bar{\lambda}_{\beta}
\ee
to obtain the following:
\be 
\mathcal{Z} &\propto& \int D\bar{\lambda}_1 D\bar{\lambda}_2 D\bar{\lambda}_3 D\bar{\lambda}_4 D\bar{\lambda}_5 D\bar{\lambda}_6 \exp \left(-N_d \sum_{\alpha} \bar{M}_{\alpha \alpha}\bar{\lambda}_{\alpha}^2+i \sqrt{N_d}\sum_{\alpha}\left(\sum_{\beta}c_{\beta}V_{\beta \alpha}\right)\bar{\lambda}_{\alpha}\right) \nn \\
\mathcal{Z} &\propto& \exp \left( -\sum_{\alpha} \frac{(\sum_{\beta}c_{\beta}V_{\beta \alpha})^2}{4 \bar{M}_{\alpha \alpha}}\right)
\ee
From this, we finally get the form 
\be 
\mathcal{Z} \propto \exp \left(-J_{eff}n^z_{1}n^z_{2}+\beta h s_1 n^z_{1}+\beta h s_2 n^z_{2} \right)
\ee
from which we can read off $J_{eff},s_1$ and $s_2$
\be 
J_{eff} &=& \sum_{\alpha} \frac{V_{1\alpha}V_{4\alpha}}{2\bar{M}_{\alpha \alpha}} \nn \\
s_1 &=& \frac{1}{2h\beta}\sum_{\alpha \beta}\frac{c_{\alpha}V_{1 \beta}V_{\alpha \beta}}{\bar{M}_{\beta \beta}} \nn \\
s_2 &=& \frac{1}{2h\beta}\sum_{\alpha \beta}\frac{c_{\alpha}V_{\alpha \beta}V_{4 \beta}}{\bar{M}_{\beta \beta}} 
\ee

\section{$T=0$ calculation of smeared spin density of orphan spin texture}
The planar pyrochlore lattice, where the four spins in an elementary unit interact equally strongly with each other (Fig~\ref{planarpyro}), presents a particularly simple case in two dimensions where the effective field theory can be solved easily because the diagonalizing matrix $U(\vec{k})$ is a $2 \times 2$ matrix (instead of being $7 \times 7$ as in SCGO). Therefore, we focus
on this tractable case to illustrate the behaviour of the smeared spin
density of a $T=0$ orphan spin texture.

\begin{figure}
{\includegraphics[width=0.3\hsize]{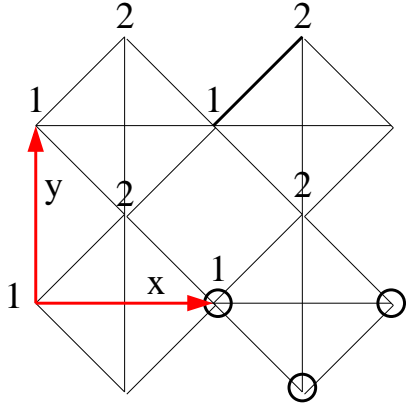}}
\caption{The planar pyrochlore lattice shown above. The lattice can be represented as a Bravais lattice with a two-point basis as shown. The lattice translation vectors $\hat{x},\hat{y}$ are also shown. Three spins have to be removed from a simplex (open circles) to produce an orphan spin.}
\label{planarpyro}
\end{figure}
{\it{Smeared total spin operator:}} As we have noted in the main text,
an orphan spin in a spin $S$ magnet induces a scale-free $1/r$ texture around it at $T=0$ in the presence of even an infinitesimal magnetic field. Formally adding up the
spin polarization at all sites in the system then leads to $S^z_{tot}=S/2$ via the argument outlined earlier. Strictly speaking, $S^z_{tot}$ is only conditionally convergent in two dimensions. However, our finite temperature results demonstrate quite clearly that assigning a total spin of $S^z_{tot} = S/2$ to the
orphan spin texture is meaningful in that the impurity susceptibility due
to a single defective simplex is asymptotically equal to the susceptibility
of a free spin $S/2$ in the low temperature limit. Moreover, as was
already highlighted in our previous Letter, this asymptotic low
temperature result remains surprisingly accurate all the way up to
temperatures of order $T \sim 0.1J$. 

It is therefore very interesting to see how the total spin converges to $S/2$ when spins only within a certain radius of the orphan spin (which drives the texture) are taken into account. To this
end, we define the following {\it{smeared}} total spin operator
\be 
S^z_{tot}(\xi) = \sum_{\vec{r}} \langle S^z(\vec{r})\rangle \exp(-(\vec{r}-\vec{r}_0)^2/\xi^2)
\ee   
where we use a Gaussian centered at $\vec{r}_0$, the orphan spin location, to regulate the conditionally convergent sum. Thus, the above smeared operator effectively ignores the contribution of the spins located at distances much greater than $\xi$ from the orphan spin---note that $\xi$ is {\em not} the thermal correlation length, but a free parameter here. It is clear that when $\xi \rightarrow 0$, $S^z_{tot}(\xi) \rightarrow S$. 

By way of illustration, we calculate $\langle S^z(\vec{r})\rangle$ and this smeared total spin $\langle S^z_{tot}(\xi)\rangle$ for the planar pyrochlore lattice in the thermodynamic limit at $T=0$ using a lattice Green's function technique. At $T=0$, we have $\sum_{\vec{r}\epsilon \XBox} S^z_{\vec{r}}=0$ and the orphan spin $S^z(\vec{r}_0)=S$ when $h \rightarrow 0$. On the dual square lattice, finding the texture $\langle S^z(\vec{r})\rangle$ is equivalent to determining the {\it current} on each bond of the lattice given that three bonds of the lattice are removed (the three vacancies around the orphan spin) and there is an input current of $S$ at $\vec{r}_0$ (the polarized orphan spin). The condition $\sum_{\vec{r}\epsilon \XBox} S^z_{\vec{r}}=0$ translates to current conservation at each site on the dual lattice. The current profile can be calculated by evaluating the lattice Green's function of the square lattice with three removed links which determines the potential on each site. The current on each bond is then just the potential difference across the bond. To calculate the lattice Green's function, we use the fact that the Green's function of the square lattice with three links removed can be easily expressed in terms of the lattice Green's function of the full square lattice (see J.~Cserti, D.~Gyula and P.~Attila, Am. J. Phys. {\bf 70} (2), 153 (2002)).

In this way, we obtain the result:
\be 
S^z_{tot}(\xi)=\frac{S}{2}+Sf(\xi)
\ee
where $f(\xi) \approx 0.16/\xi^2$ when $\xi \gg 1$; this
behaviour is displayed in Fig~\ref{smearedcharge}.
\begin{figure}
{\includegraphics[width=\hsize]{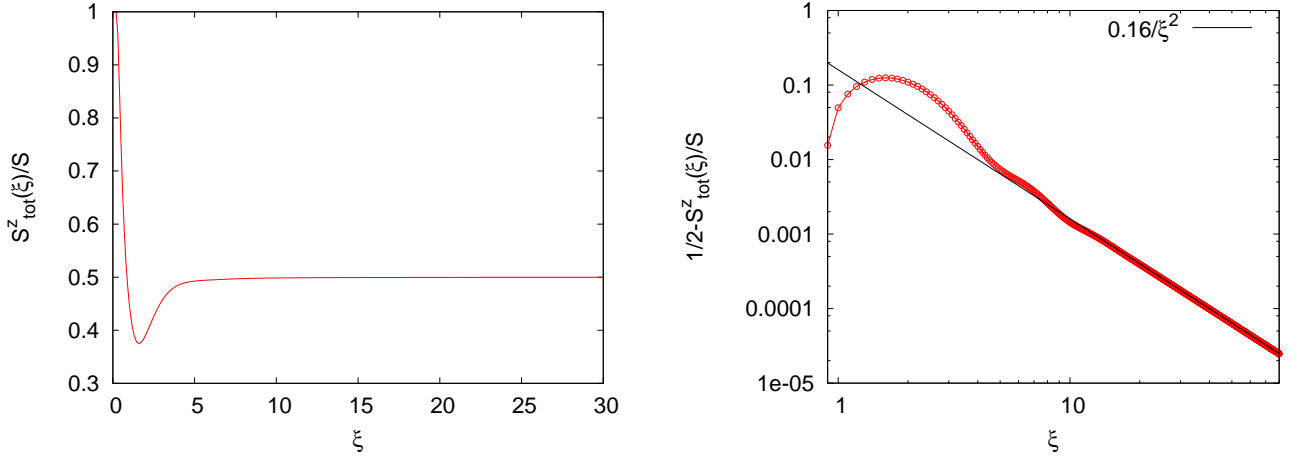}}
\caption{Exact calculation of the smeared total spin operator $S^z_{tot}(\xi)$ for the planar pyrochlore lattice at $T=0$; values on the $y$ axis are
measured in units of $S$. The solution shows that the spin density is mostly concentrated in the neighborhood of the orphan spin and decays rapidly away from it.}
\label{smearedcharge}
\end{figure}
Thus, $S^z_{tot}(\xi)$ approaches $S/2$ quickly (Fig~\ref{smearedcharge}) and it is already within $1\%$ of its $\xi \rightarrow \infty$ value when $\xi \sim 6$ for the planar pyrochlore lattice. In other words, the spin density, which peaks at the orphan spin, is concentrated close to it and falls rapidly as the distance from the orphan increases. And clearly, it is this rapid fall-off
that is at the root of the surprising robustness of the orphan physics
all the way up to temperatures of order $T \sim 0.1JS^2$.

\end{widetext}


\begin{thebibliography}{999}
\bibitem{Moessner_Ramirez} R.~Moessner, and A.~P.~Ramirez,  Phys. Today {\bf 59}, 24 (2006).   
  
\bibitem{Rajaraman_fractionalization} R. Rajaraman,  {\em Quantum (Un)speakables} 383-399, R.A.Bertlemann and A.Zeilinger (Editors), Springer-Verlag (Berlin) (2002). 


\bibitem{Castelnovo_Moessner_Sondhi}
C.~Castelnovo, R.~Moessner, and S.~L.~Sondhi, Nature {\bf 451}, 42 (2008).

\bibitem{HusKrauMoesSond} D.~A.~Huse, W.~Krauth, R.~Moessner, and S.~L.~Sondhi, Phys. Rev. Lett. {\bf 91}, 167004 (2003).

\bibitem{Hermele_Balents_Fisher} M.~Hermele, M.~P.~A.~Fisher, and L.~Balents,
Phys. Rev. B {\bf 69}, 064404 (2004).

\bibitem{Banerjee_etal} A.~Banerjee, S.~Isakov, K.~Damle, and Y.~B.~Kim,
Phys. Rev. Lett. {\bf 100}, 047208 (2008).

\bibitem{Shannon_etal} N.~Shannon {\em et. al.}, arXiv: 11054196 (unpublished).



\bibitem{ober1} X.~Obradors {\em et. al.}, Solid State Commun. {\bf 65}, 189 (1988).

\bibitem{ober2}  B.~Martinez, A.~Labarta, R.~Rodriguez-Sola, and X.~Obradors,
Phys. Rev. B {\bf 50} 15779 (1994).

\bibitem{ramirez} A.~P.~Ramirez, G.~P.~Espinosa, and A.~S.~Cooper, Phys. Rev. Lett. {\bf 64}, 2070 (1990).

\bibitem{kerenMUSR} Y.~J.~Uemura {\em et. al.}, Phys. Rev. Lett. {\bf 73}, 3306
(1994).

\bibitem{schifferSUS} P.~Schiffer and I.~Daruka, Phys. Rev. B {\bf 56},
13712 (1997).


\bibitem{Limot_etal_prb} L.~Limot {\em et. al.}, Phys. Rev. B {\bf 65}, 144447 (2002).

\bibitem{Sen_Damle_Moessner_PRL} A.~Sen, K.~Damle, and R.~Moessner,
Phys. Rev. Lett. {\bf 106}, 127203 (2011).


\bibitem{Moessner_berlinsky} R.~Moessner and A.~J.~Berlinsky, Phys. Rev. Lett.
{\bf 83}, 3293 (1999).

\bibitem{Henley_2000} C.~L.~Henley, Can. J. Phys. {\bf 79}, 1307 (2001).



\bibitem{Henley_effectivetheory} C.~L.~Henley in {\em Annual Review of Condensed Matter Physics} Vol. 1,  179-210 (2010). 

\bibitem{Garanin_Canals}D.~A.~Garanin
and B.~Canals, Phys. Rev. B {\bf 59}, 443 (1999).

\bibitem{Isakov_Moessner_Sondhi} S.~V.~Isakov, K.~Gregor, R.~Moessner,
and S.~L.~Sondhi, Phys. Rev. Lett. {\bf 93}, 167204 (2004).

\bibitem{Lee_Young} L. W. Lee and A. P. Young,
Phys. Rev. B {\bf 76,} 024405 (2007).

\bibitem{chalker1} R.~Moessner and J.~T.~Chalker, Phys. Rev. Lett. {\bf 80}, 2929 (1998); Phys. Rev. B {\bf 58}, 12049 (1998).

\bibitem{chalker2} P.~H.~Conlon and J.~T.~Chalker, Phys. Rev. Lett. {\bf 102}, 237206 (2009).

\bibitem{Keren} A.~Keren {\em et. al.}, Phys. Rev. Lett.  {\bf 72}, 3254 (1994).

\bibitem{Villian} J.~Villian, Zeitshrift fur Physik B Cond. Matt.  {\bf 33},
31, (1979).

\bibitem{Damle_etal} K.~Damle {\em et. al.}, work in progress (unpublished).







\end{thebibliography}
\end{document}